\documentclass[10pt,twocolumn]{article}

\usepackage[pagenumbers]{cvpr}      % To produce the REVIEW version
\usepackage{graphicx}    
\usepackage{float} 
\usepackage{caption}

\graphicspath{ {./fig2/} {./fig3/}{./fig4/}{./fig5/}{./fig6/}{./fig7/}{./fig8/}}
\usepackage[table,xcdraw]{xcolor}
\title{PSAvatar: A Point-based Shape Model for Real-Time Head Avatar Animation with 3D Gaussian Splatting}

\author{Zhongyuan Zhao$^{1,2}$, Zhenyu Bao$^{1,2}$, Qing Li$^1$, Guoping Qiu$^{3,4}$, Kanglin Liu $^1$
	\\
$^1$ Pengcheng Laboratory 
$^2$ Peking University 
$^3$ University of Nottingham
$^4$ Shenzhen University
}

\begin{document}
\twocolumn[{%
	\renewcommand\twocolumn[1][]{#1}%
	\maketitle
	\includegraphics[width=.98\linewidth]{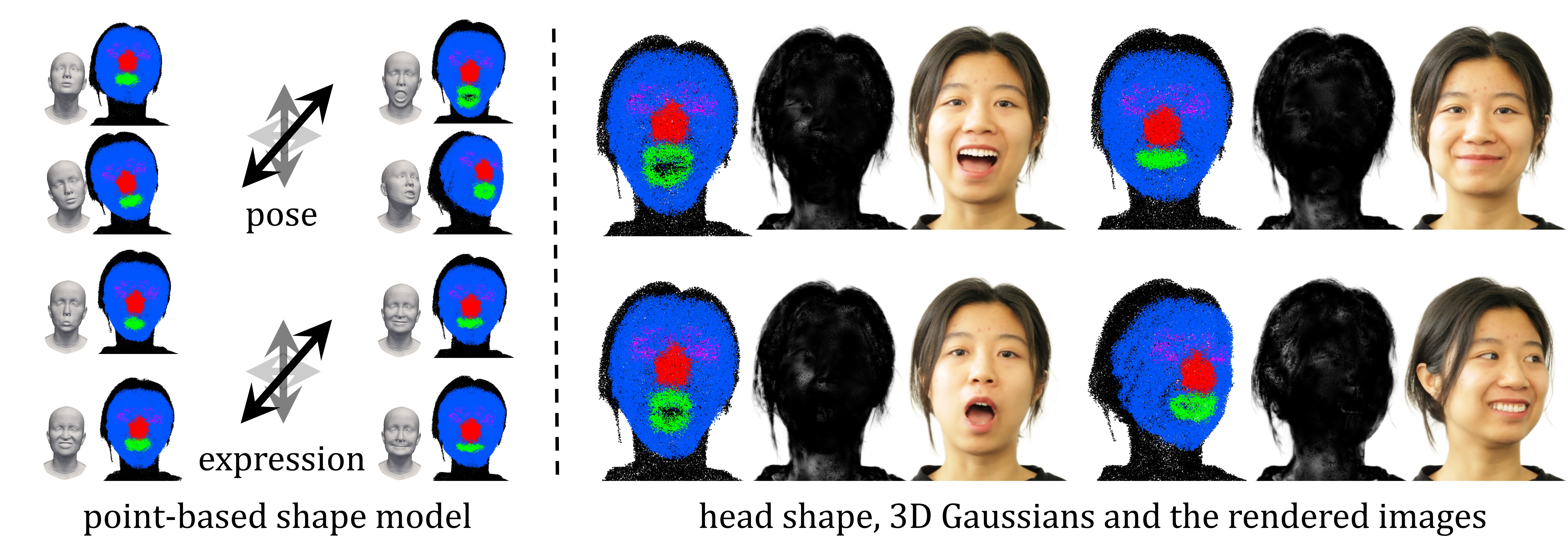}	
	% \vspace{-2em}
	\captionof{figure}{PSAvatar learns the shape with pose and expression variations based on a point-based shape model, and employs 3D Gaussian for fine detail representation and efficient rendering. Given monocular portrait videos, PSAvatar can create head avatars that enable real-time ($\ge$ 25 fps at 512 $\times$ 512 resolution) and high-fidelity rendering.}
	\vspace{2em}
	\label{fig:teaser}
}]

\begin{abstract}
	Despite much progress, achieving real-time high-fidelity head avatar animation is still difficult and existing methods have to trade-off between speed and quality. 3DMM based methods often fail to model non-facial structures such as eyeglasses and hairstyles, while neural implicit models suffer from deformation inflexibility and rendering inefficiency.
	Although 3D Gaussian has been demonstrated to possess promising capability for geometry representation and radiance field reconstruction, applying 3D Gaussian in head avatar creation remains a major challenge since it is difficult for 3D Gaussian to model the head shape variations caused by changing poses and expressions. In this paper, we introduce PSAvatar, a novel framework for animatable head avatar creation that utilizes discrete geometric primitive to create a Point-based Shape Model (PSM) and employs 3D Gaussian for fine detail representation and high fidelity rendering. 
	The PSM uses points instead of meshes for 3D representation to achieve enhanced representation flexibility. 
	%Based on the inherited capability of the parametric morphable face model FLAME,  the PSM is competent to use points instead of meshes for 3D representation to achieve enhanced representation flexibility. 
	Specifically, PSM first converts the FLAME mesh to points by sampling on the surfaces as well as off the meshes to enable the reconstruction of not only surface-like structures but also complex geometries such as eyeglasses and hairstyles. By aligning these points with the head shape in an analysis-by-synthesis manner, the PSM makes it possible to utilize 3D Gaussian for fine detail representation and appearance modeling, thus enabling the creation of high-fidelity avatars. We show that PSAvatar can reconstruct high-fidelity head avatars of a variety of subjects and the avatars can be animated in real-time ($\ge$ 25 fps at a resolution of 512 $\times$  512 )\footnote{Test is conducted based on Nvidia RTX 3090}.
\end{abstract}

\section{Introduction}

Creating animatable head avatars has wide applications and has attracted extensive interests in academia and industries. %Recent methods for creation of head avatars from portrait videos rely on either 
Many methods based on explicit representations, \textit{e.g.}, 3D morphable models (3DMMs) \cite{Blanz99, Li17}, points \cite{Zheng23, Xu22} and more recently 3D Gaussian \cite{Kerbl23, Qian23, Chen23}), and neural implicit representations, \textit{e.g.}, Neural Radiance Field (NeRF) \cite{Mildenhall21, Gafni21, Zielonka23} and signed distance function (SDF) \cite{Yariv20, Zheng22}), have been developed in recent years. Whilst these methods have achieved very impressive results, there are still many unsolved problems. 

3DMM-based methods allow efficient rasterization and inherently generalize to unseen deformations, but are limited by a priori-fixed topology and surface-like geometries, making them less suitable for modeling individuals with eyeglasses or complex hairstyles \cite{Chen23, Qian23}. Whilst neural implicit representations outperform 3DMM-based methods in capturing hair strands and eyeglasses \cite{Zheng22, Duan23}, they are computationally extremely demanding \cite{Grassal22}. %. However, rendering a pixel via neural implicit representations requires sampling numerous points along camera rays, significantly affecting the training and rendering efficiency \cite{Grassal22}. 
%Besides, due to the absence of the prior deformable topology, 
Furthermore, neural implicit representations need the deformer network or similar techniques to bridge the gap between the canonical and deformed spaces, making it challenging to achieve high deformation accuracy.

In contrast to neural implicit representations, both point and 3D Gaussian representations can be rendered efficiently with a splatting-based rasterization \cite{Zheng23, Chen23, Qian23}, and both are considerably more flexible than 3DMMs in representing complex volumetric structures, \textit{e.g.}, eyeglass, hair strands, \textit{etc.}. PointAvatar \cite{Zheng23} initializes with a sparse point cloud randomly sampled on a sphere and periodically upsamples the point cloud by adding noises. The position of the points are updated to match the target geometry via backwards gradients. Points are rotation-invariant and isotropically scaled, making them easy to control. In comparison, 3D Gaussians can be rotated and scaled, making them more flexible than points for 3D representation. In order to achieve consistent 3D representations, 3D Gaussian rely on carefully designed controlling strategy. In GaussianAvatar \cite{Qian23}, each triangle of the mesh is initialized with a 3D Gaussian, and the positional gradient is utilized to move and periodically densify the Gaussian splats. 
A major difficulty in applying 3D Gaussian to head avatar creation is modeling the head shape variations caused by changing poses and expressions.

In this paper, we introduce PSAvatar, a novel framework for animatable head avatar creation that utilizes discrete geometric primitive to create a point-based shape model and employs 3D Gaussian for fine detail representation and high fidelity rendering. The PSM relies on points instead of meshes for 3D representation to achieve enhanced representation flexibility. Specifically, PSM converts the FLAME mesh to points by uniformly sampling points on the surface of the mesh. However, FLAME is incapable of representing individuals with eyeglasses or complex hairstyles. To address this, PSM samples points off the FLAME mesh to enhance the representation flexibility. 
PSM splats the points onto screen and minimizes the difference between the rendered and ground truth images. After removing the invisible points, the remaining points are then aligned with the head shape.
%During the alignment stage, SMSM utilizes point splats for 3D representation as points are with less trainable parameters than that of Gaussian splats, and are easy to train and converge quickly to approximate the head shape.
PSAvatar models the appearance by employing 3D Gaussian in combination with the PSM to reconstruct the underlying radiance field and to achieve high-fidelity rendering. Our contributions are as follows: 

% itemize
\begin{itemize}
	\item We present PSAvatar, a method for creating animatable head avatars using a point-based shape model for shape modeling and employing 3D Gaussian for fine detail representation and appearance modeling.
	\item We have developed a Point-based Shape Model for 3D head representation that is capable of modeling facial shapes with pose and expression variations and capturing complex volumetric structures \textit{e.g.,} hair strands, glasses, \textit{etc.}.
	\item We show that PSAvatar can reconstruct high-fidelity head avatars of a variety of subjects and the avatars can be animated in real-time ($\ge$ 25 fps at 512 $\times$ 512 resolution). % by the morphable model parameters.  and implement real-time animation driven by  the morphable parameters.
\end{itemize}

% Head 1
\section{Related Work}
%\noindent{\textbf{Head Avatar Creation with Implicit Models}}
\noindent{\textbf{Head Avatar Creation with Implicit Models}}
%Implicit models reconstruct the face by neural radiance field in combination with volumetric rendering or using implicit surface functions (\textit{e.g.}, signed distance functions) along with the shader networks to represent the geometry and reconstruct the appearance, respectively. 
%In order to create an animatable head avatar, 
Implicit models reconstruct the face by neural radiance field in combination with volumetric rendering or using implicit surface functions (\textit{e.g.}, signed distance functions).
A popular approach of creating animatable head avatar is to condition the NeRF on low-dimensional facial model parameters such as expression, pose and camera setting \cite{Gafni21, Wang21, Gao21}. 
NeRFBlendshape \cite{Gao22} models the dynamic NeRF by linear combinations of multiple NeRF basis one-to-one corresponding to semantic blendshape coefficients.
%NeRFBlendshape \cite{Gao22} constructs personalized semantic facial model based on multi-level voxel fields. 
%Even though attempts have been made towards solving the geometric and temporal inconsistencies, 
Despite achieving impressive performances, such an approach could either fail to disentangle pose and expression or fail to generalize well to novel poses and expressions \cite{Egger20, Gafni21, Kellnhofer21, Ramon21, Su21, Wang21:A, Wang21:B}.
Another paradigm is to establish the target head model in the canonical space %that can be deformed to 
and synthesize the dynamics by deformation \cite{Kocabas23, Lei23, Ye23}. 
INSTA \cite{Zielonka23} deforms the query points from the observation space to the canonical space by using the bounding volume hierarchy (BVH) and employs InstantNGP to accelerate rendering. 
IMAvatar \cite{Zheng22} represents the deformation fields via learned expression blendshapes and %linear blend skinning weights, and 
solves for the mapping from the observed space to the canonical space via iterative root-finding. 
AvatarMAV \cite{Xu23} defines motion-aware neural voxels, and models deformations via blending a set of voxel grid motion bases according to an input 3DMM expression vector.
In addition, a variety of powerful techniques such as triplane \cite{Zhao23}, Kplane \cite{Fridovich-Keil23}, deformable multi-layer meshes \cite{Duan23} have been utilized in head avatar creation to improve training efficiency and rendering quality.

\begin{figure*}[!h]
	\centering
	\includegraphics[width=7in]{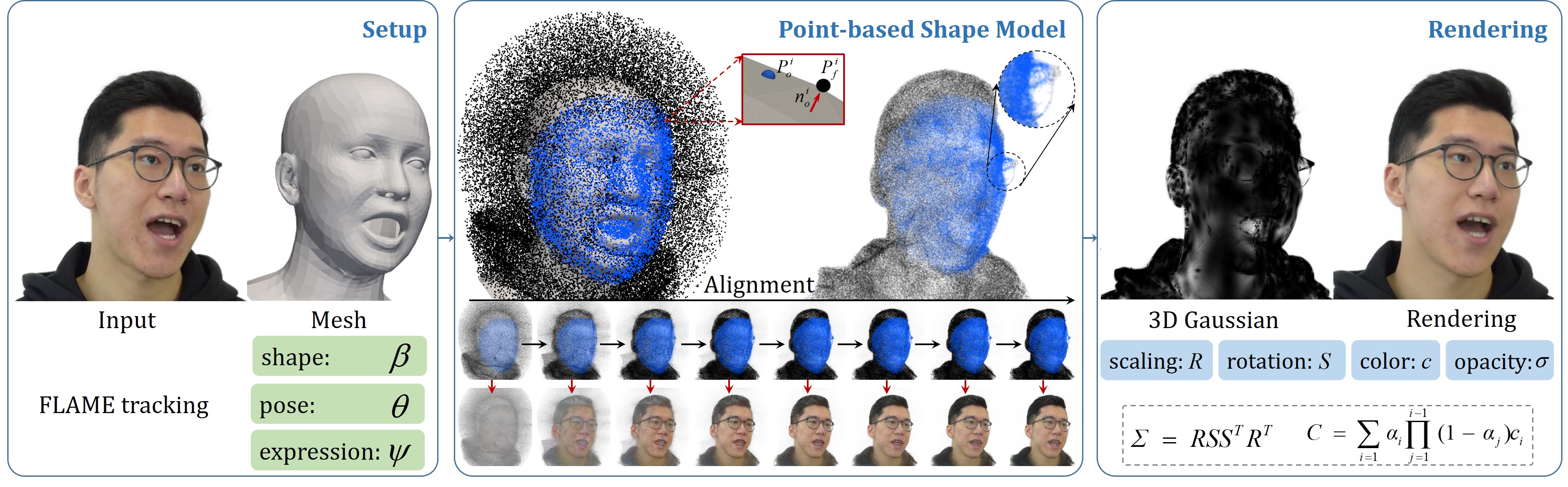}
	\caption{Overview. Given a monocular portrait video, we conduct FLAME tracking to obtain the parameters. The point-based shape model (PSM) first converts the FLAME mesh to points. It performs sampling on the surfaces (blue points) and additionally generates samples off the meshes by offsetting the samples on the meshes along their normal directions (black points). These points are then aligned with the head shape in an analysis-by-synthesis manner. The inclusion of points on meshes and off meshes enables the PSM to reconstruct not only surface-like structures but also complex geometries that are beyond the capability of 3DMMs. Combining the PSM with 3D Gaussian allows the reconstruction of the radiance field for efficient rendering.}
	\label{fig:two}
\end{figure*}

\noindent{\textbf{Head Avatar Creation with Explicit Models}}
The seminal work of 3D Morphable Model (3DMM) \cite{Blanz99} uses principal component analysis (PCA) to model facial appearance and geometry on a low-dimensional linear subspace.
3DMM and its variants \cite{Li17,Paysan09,Cao13,Gerig18,Yao21} have been widely applied in optimization-based and deep learning-based head avatar creation \cite{Gecer19, Tewari18, Deng19, Dib21}.
Neural Head Avatar \cite{Grassal22} employs neural networks to predict
vertex offsets and textures, enabling the extrapolation to unseen facial expressions. ROME \cite{Khakhulin22} estimates a person-specific head mesh and the associated neural texture to enhance local photometric and geometric details.
3DMM-based methods produce geometrically consistent avatars that can be easily controlled, however, they are limited to craniofacial structures and can fail to represent hair and glasses. % \textit{etc.}, greatly affecting the rendering quality.
To address this, PointAvatar \cite{Zheng23} explores point-based geometry representation with differential point splatting, allowing for high quality rendering and representation of hair and eyeglasses. 
%Points are rotation-invariant and isotropically scaled which makes them easy to handle, however, point representation lacks flexibility. 
Although point-based representations are easy to handle, they lack flexibility. In comparison, 3D Gaussian \cite{Kerbl23} offers improved flexibility. % in terms of scale and rotation. 
MonoGaussianAvatar \cite{Chen23} replaces the point cloud in PointAvatar with Gaussian points to improve representation flexibility and rendering quality.  %improving the representation ability and rendering quality.
%As for GaussianAvatars \cite{Qian23}, each triangle of the mesh is paired with a 3D Gaussian in the initialization stage, and the densification and pruning strategies are introduced for sufficient representation of the geometry. Besides, the binding inheritance ensures that the 3D Gaussian translates and rotates with the triangle, allowing for precise animation control via the underlying parametric model.
%
% added by zy
SplattingAvatar \cite{Shao24} employs a trainable embedding strategy for Gaussian-Mesh association.
%it uses a walking strategy to update the embedding and enhance rendering capabilities. 
FlashAvatar \cite{Xiang2023} improves the geometric representation ability by maintaining a 3D Gaussian field in 2D UV space.
%and then embedding the Gaussians into dynamic FLAME mesh surfaces.
%
GaussianAvatars \cite{Qian23} pairs a triangle of the mesh with a 3D Gaussian and introduces densification and pruning strategies for sufficiently representing the geometry. In addition, it uses binding inheritance to ensure that the 3D Gaussian translates and rotates with the triangle to enable precise animation control via the underlying parametric model. 

Both point and 3D Gaussian rely on the controlling strategy for achieving consistent 3D representation.
PointAvatar \cite{Zheng23} initializes sparse points by randomly sampling on a sphere, and updates their positions to approximate the coarse shape. Besides, PointAvatar introduces a deformer network to bridge the gap between the canonical space and the deformed space. To model details, it periodically densifies points by adding noises.
3D Gaussians are considerably more flexible than points in 3D representation but are more difficult to control.
% added by zy
SplattingAvatar \cite{Shao24} utilizes the learnable barycentric coordinate and a displacement along the interpolated normal to locate the correspondences between 3D Gaussian and the underlying mesh. FlashAvatar \cite{Xiang2023} attaches 3D Gaussians to the head mesh based on the fixed UV mapping and learns dynamic offsets to improve the the representation capacities. 
In GaussianAvatars \cite{Qian23}, each triangle of the mesh is initialized with a 3D Gaussian. For each 3D Gaussian with a large positional gradient, GaussianAvatars splits it into two smaller ones if it is large or clone it if it is small. The newly generated 3D Gaussian is bound to the same triangle as the old one to enable binding inheritance during densification. A major difficulty in applying 3D Gaussian to head avatar creation is modeling the head shape variations caused by changing poses and expressions. Our new PSAvatar uses a point-based shape model to capture the head dynamics and successfully achieve real-time head avatar animation.

%However, these operations make the process of reconstructing the geometry very slow. Our new PSAvatar uses a point-based morphable shape model to  speed up this process and successfully achieve real-time head avatar animation. %However, these operations make the process of reconstructing the geometry very slow. Our new PSAvatar uses a point-based morphable shape model to greatly speed up this process and makes it possible to create head avatars in real-time.  %the initialization and densification %strategies result in slow convergence for points and Gaussians to reconstruct the geometry. 

\section{Method}
Fig. \ref{fig:two} shows the schematic of PSAvatar. The objective is to reconstruct an animatable head avatar with a monocular portrait video of a subject performing diverse expressions and poses. To achieve this, PSAvatar introduces a point-based shape model (PSM) for 3D representation to model pose and expression dependent shape variations (see section \ref{MSM-S}), and models the appearance by combining the PSM and 3D Gaussian (see section \ref{Rendering}). 
%PSAvatar reconstructs the radiance field in an analysis-by-synthesis fashion, \textit{i.e,}, splats are rendered into images via a differentiable tile rasterizer, that are then supervised by the portrait video frames.
\begin{figure*}	[!t]
	\centering
	\subfloat[]{
		\begin{minipage}[b]{0.24\linewidth}		
			\includegraphics[width=0.45\linewidth]{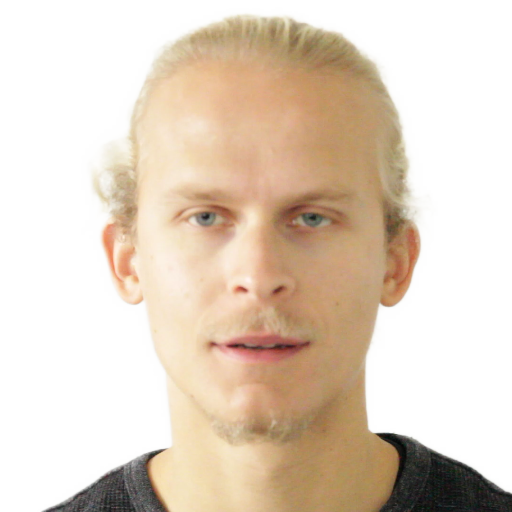}
			\includegraphics[width=0.45\linewidth]{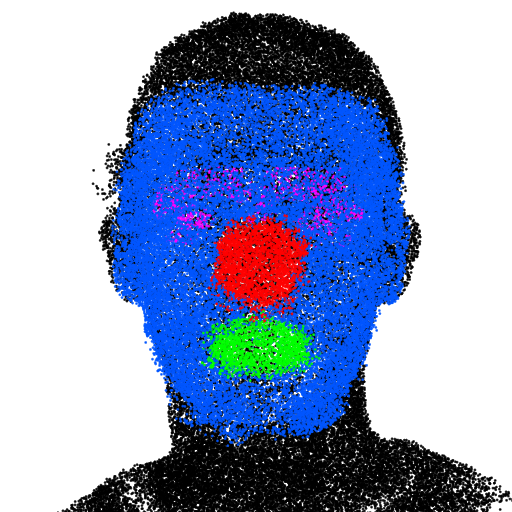}			
		\end{minipage}
	}
	\subfloat[]{
		\begin{minipage}[b]{0.24\linewidth}
			\includegraphics[width=0.45\linewidth]{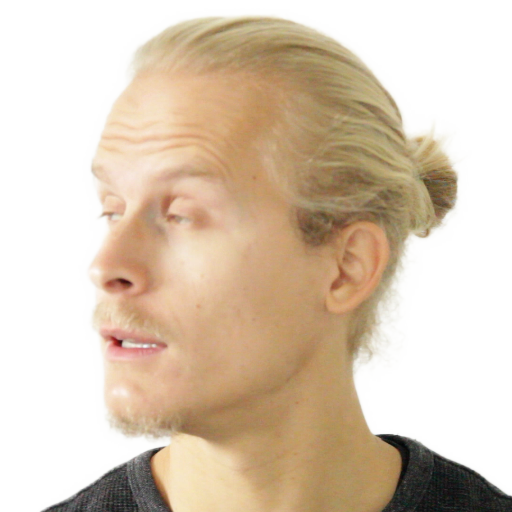}
			\includegraphics[width=0.45\linewidth]{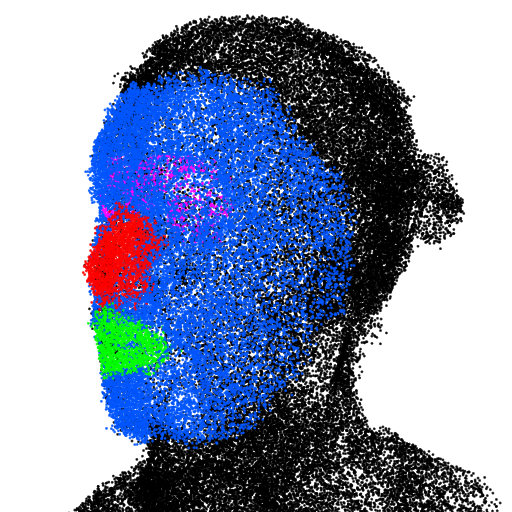}			
		\end{minipage}
	}
	\subfloat[]{
		\begin{minipage}[b]{0.24\linewidth}
			\includegraphics[width=0.45\linewidth]{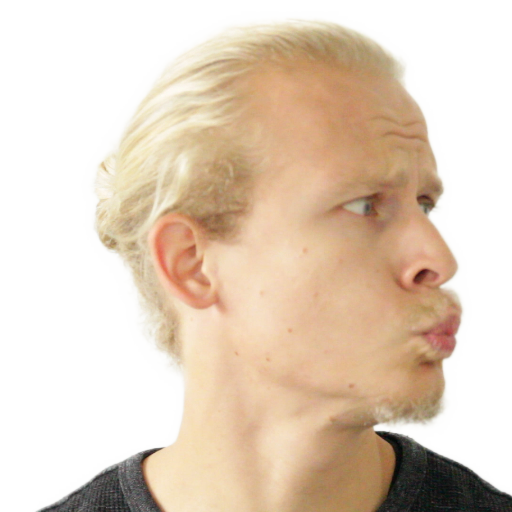}
			\includegraphics[width=0.45\linewidth]{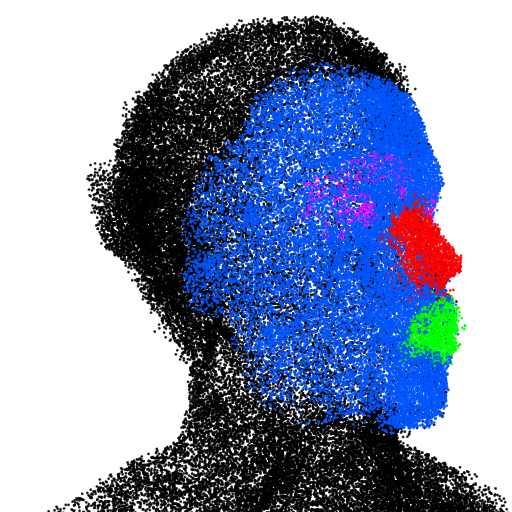}		
		\end{minipage}
	}
	\subfloat[]{
		\begin{minipage}[b]{0.24\linewidth}
			\includegraphics[width=0.45\linewidth]{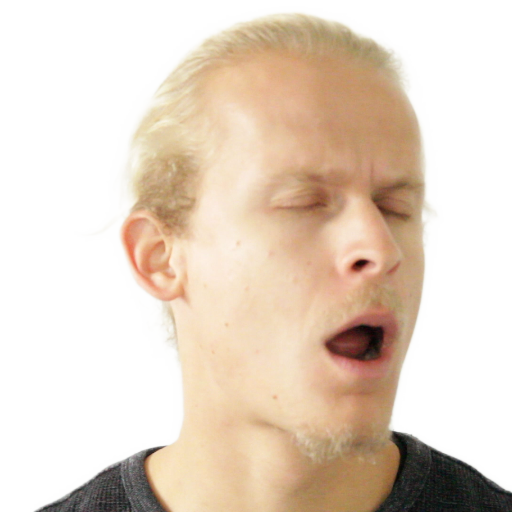}
			\includegraphics[width=0.45\linewidth]{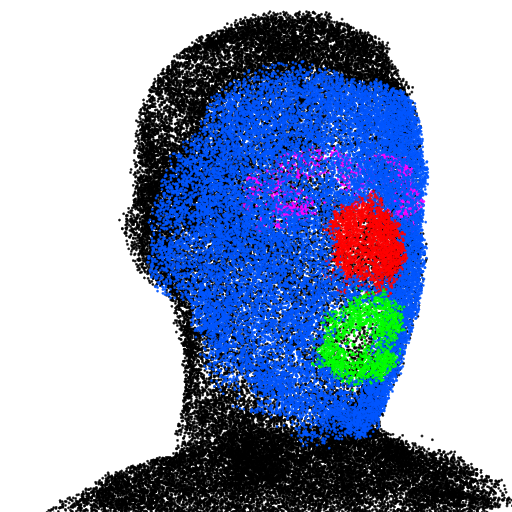}		
		\end{minipage}
	}		
	\caption{Shape variations for given poses and expressions. 
		The reference images (taken from subject 2) on the left provide the pose and expression parameters, and the	Point-based Shape Model (PSM) can warp the points in a way that is consistent with the reference, \textit{i.e.} the reference person turns his head around, the points follow the movements. Blue and black represent the points on and off the mesh respectively. To visualize the shape variation in a better way, points sampled based on the eye, nose and mouth regions are colored with pink, red and green, respectively.}
	\label{fig:three}	
\end{figure*}

\subsection{Preliminary}
Point and 3D Gaussian utilize discrete primitives for geometry representation. Points are parameterized by the radius $r$, the opacity $\sigma$ and the color $c$. A 3D Gaussian is defined by a covariance matrix $\Sigma $ centered at a point (mean) $\mu$ \cite{Kerbl23}:
\begin{equation}
	\label{eqn:01}
	G(x)=e^{-\frac{1}{2}(x-\mu)^T\Sigma(x-\mu)}
\end{equation}
To guarantee that $\Sigma$ is physically meaningful, the covariance matrix is constructed by a parametric ellipse with a scaling matrix $S$ and a rotation matrix $R$:
\begin{equation}
	\label{eqn:02}
	\Sigma = RSS^TR^T
\end{equation}
Identical to that in \cite{Kerbl23}, the scaling and rotation matrix are represented by a scaling vector $s \in R^3$ and  a quaternion $q \in R^4$, respectively.
$s$ and $q$ can be trivially converted to their respective matrices, and they can be combined to make sure that the normalised $q$ is a valid unit quaternion. % and combined %, making sure to normalize $q$ to obtain a valid unit quaternion.

Both points and 3D Gaussian can be rendered via a differentiable splatting-based rasterizer, and the color $C$ of a pixel is computed by alpha compositing:
\begin{equation}
	\label{eqn:03}
	C = \sum_{i=1}\omega_i c_i =\sum_{i=1}\alpha_i\prod_{j=1}^{i-1} (1-\alpha _j)c_i
\end{equation}
where $\omega_i$ is the weight for alpha compositing, $c_i$ is the color of each point or 3D Gaussian and $\alpha$ is the blending weight. For points, $\alpha$ is calculated as $\alpha = \sigma (1-d^2/r^2)$, where $d$ is the distance from the point center to the pixel center. For Gaussians, $\alpha$ is given by evaluating the 2D projection of the 3D Gaussian multiplied by a per-point opacity $\sigma$.

\subsection{Point-based Shape Model} \label{MSM-S}

Given shape, pose and expression components, FLAME \cite{Li17} can produce morphologically realistic faces in a convenient and effective way. This motivates us to build a point-based shape model (PSM) on FLAME to inherit its morphable capability. Since we focus on human facial avatars, we specifically model the pose and expression dependent shape variations and simplify the FLAME model $M(\theta, \psi )$:
\begin{equation}
	\label{eqn:04}
	M(\theta, \psi  )=W(T_P(\theta, \psi ), J(\psi ), \theta, \mathcal{W})
\end{equation}
where $\theta$ and $\psi$ denote the pose and expression parameters respectively. $W(\cdot)$ and $J(\cdot)$ define the standard skinning function and the joint regressor respectively. $\mathcal{W}$ represents the per-vertex skinning weights for smooth blending, and $T_P$ denotes the template mesh with pose and expression offsets, defined as:
\begin{equation}
	\label{eqn:05}
	T_P(\theta, \psi ) = \bar{T} + B_P(\theta; \mathcal{P}) + B_E(\psi ; \mathcal{E}) + \mathcal{G} (\theta, \psi)
\end{equation}
where $\bar{T}$ is the personalized template, $B_P$ and $B_E$ model the corrective pose and expression blendshapes, respectively. $\mathcal{P}$ and $\mathcal{E}$ denote the pose and expression basis, respectively.
To model the inconsistency between FLAME and the head geometry, $ \mathcal{G} (\theta, \psi)$ is introduced as the per-vertex geometry correction:
\begin{equation}
	\label{eqn:06}
	\mathcal{G} (\theta, \psi) = B_P(\theta; \mathcal{P}^{'}) + B_E(\psi ; \mathcal{E}^{'})
\end{equation}
where $\mathcal{P}^{'}$ and $\mathcal{E}^{'}$ are learned pose and expression blendshape basis, respectively.

%For the limitation of the craniofacial structure, the synthesized faces by 
Because FLAME is incapable of modeling hair strands or eyeglasses, PSAvatar addresses this limitation by introducing the point-based shape model (PSM) which utilizes points instead of meshes to enhance the representation flexibility. Specifically, we first convert the FLAME mesh to points by uniformly sampling points on the surface of the mesh with a probability that is proportional to the face area:
\begin{equation}
	\label{eqn:07}
	P_{o}^{i} = \sum_{j=0}^{2}\alpha_j^i v_j^{i}
\end{equation}
where the superscript $i$ denotes the $i$-th mesh, the subscript $o$ represents sampling on the mesh. 
$\alpha_j^i$ and $v_j^{i}$ with $j=$\{0, 1, 2\} are the barycentric coordinate and the vertex of the $i$-th mesh, respectively.
Since FLAME excels in modeling the facial dynamics, sampling is only conducted on the facial region. % (see Supp.Mat. for details).
In addition, sampling is conducted off the FLAME meshes to capture complex structures ignored by the FLAME model.
This is achieved by offsetting the sample on the mesh along its normal:
\begin{equation}
	\label{eqn:08}
	P_{f}^{i} = P_{o}^{i} + L_{f}^{i} \cdot n_{o}^{i}
\end{equation}
where the subscript $f$ represents sampling off the mesh. $L_{f}^{i}$ is a random offset from a uniform distribution $\left [0, L_{max}  \right ] $, where $L_{max}$ is a hyperparameter, empirically taken as 0.30  for covering the entire head as much as possible. $n_{o}^{i}$ is the normal on $P_{o}^{i}$, calculated by:
\begin{equation}
	\label{eqn:09}
	n_{o}^{i} = \sum_{j=0}^{2} \alpha_j^i n_j^{i}
\end{equation}
where  $n_j^{i}$ with $j=$\{0, 1, 2\} are the vertex normal of the $i$-th mesh.

Samples on the mesh are parameterized by the face index $i$, the barycentric coordinates, the opacity $\sigma$ and the color $c$. While samples off the mesh carry one additional parameter $L_{f}^{i}$.
During shape acquisition, the color $c$ is modeled by the RGB value for simplification.
Such a parameterization guarantees samples across diverse poses and expressions are in one to one (or point to point) correspondence, thus enabling the PSM to be morphable. % capability of PMSM.

PSM aligns the points and the target head in an analysis-by-synthesis manner, \textit{i.e.},
all the samples are splatted onto the screen via the tile rasterizer in Equation (\ref{eqn:03}), and the difference between the rendered image and the input are minimized. % for optimizing the color and opacity (see Supp.Mat. for details). 
Samples with visibility below a predefined threshold are removed.
For shape acquisition, point splatting instead of Gaussian splatting is applied as point has fewer parameters than Gaussian thus converging faster to model the shape (see section \ref{Ablation}).
As shown in Fig. \ref{fig:three} and the supplementary video sequences, the resulting shape model can represent the head geometry including the hair, and can be morphed with given poses and expressions. 

\begin{figure}[!t]
	\centering
	\subfloat[shape]{
		%			\rotatebox[origin=c]{90}{ \normalsize{NeuralRGBD NeuralRGBD} \hspace{-2.2cm}}		
		\begin{minipage}[b]{0.23\linewidth}				
			\includegraphics[width=\linewidth]{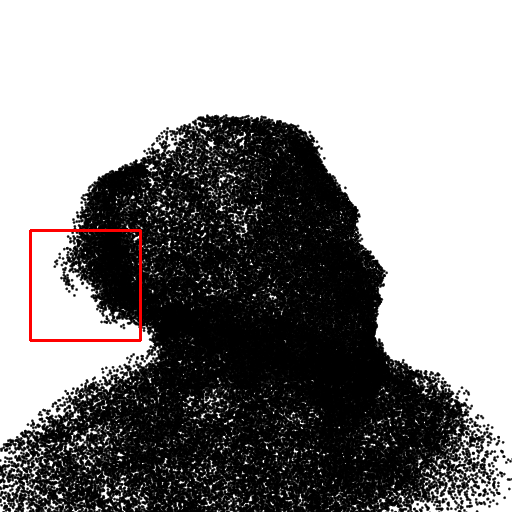}
			\includegraphics[width=\linewidth]{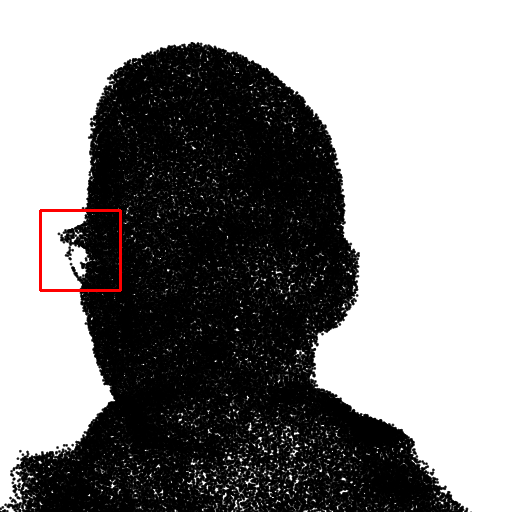}
			\includegraphics[width=\linewidth]{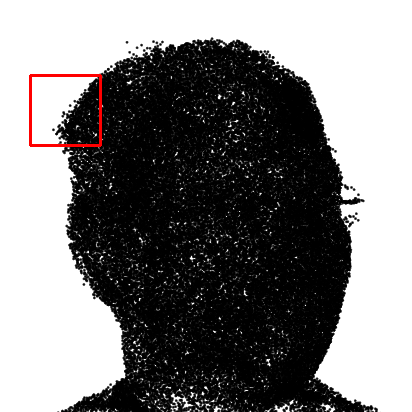}
		\end{minipage}
	}
	\subfloat[3D Gaussian]{
		\begin{minipage}[b]{0.23\linewidth}		
			\includegraphics[width=\linewidth]{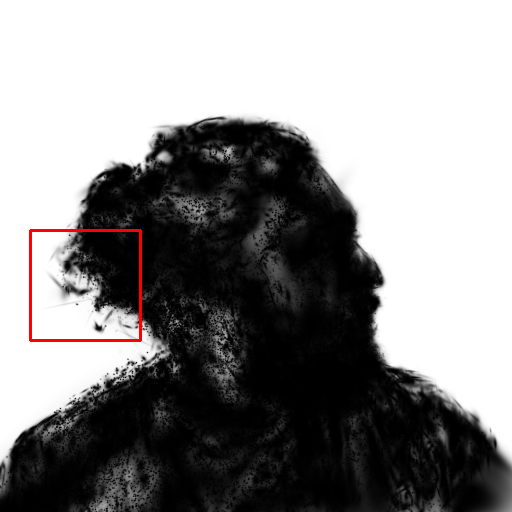}
			\includegraphics[width=\linewidth]{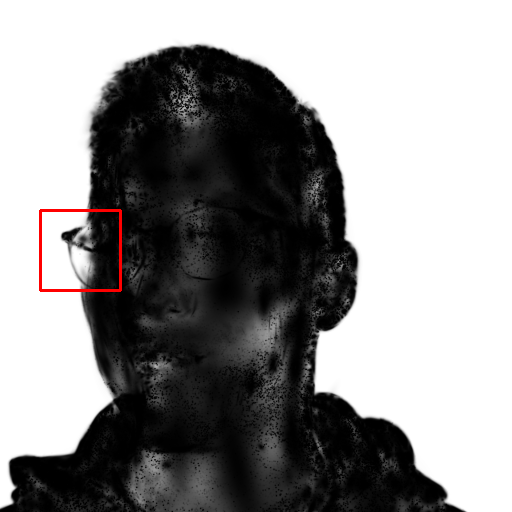}
			\includegraphics[width=\linewidth]{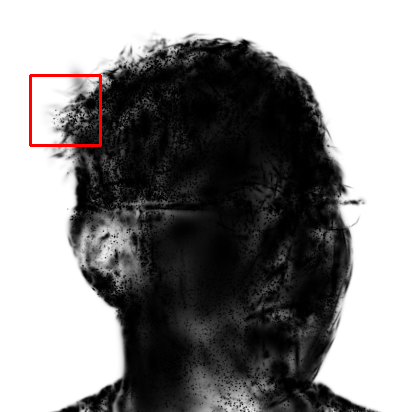}
		\end{minipage}
	}
	%	\subfloat[point splats]{
		%		\begin{minipage}[b]{0.15\linewidth}		
			%			\includegraphics[width=\linewidth]{fig4_42_2-ps.png}
			%			\includegraphics[width=\linewidth]{fig4_64_3-ps.png}
			%			\includegraphics[width=\linewidth]{fig4_45_4-ps.png}
			%			\includegraphics[width=\linewidth]{fig4_0_5-ps.png}
			%			\includegraphics[width=\linewidth]{fig4_31_6-ps.png}
			%		\end{minipage}
		%	}
	\subfloat[rendered]{
		\begin{minipage}[b]{0.23\linewidth}		
			\includegraphics[width=\linewidth]{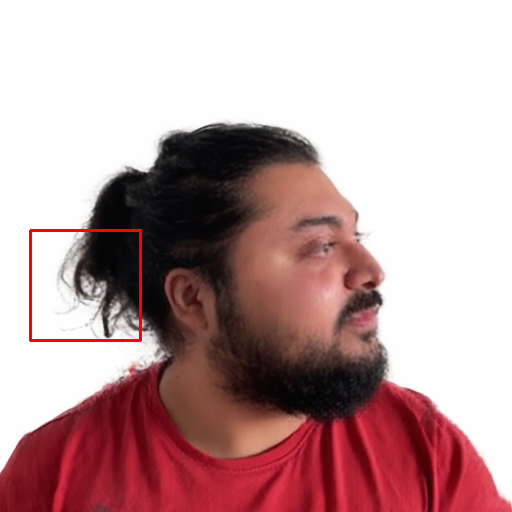}
			\includegraphics[width=\linewidth]{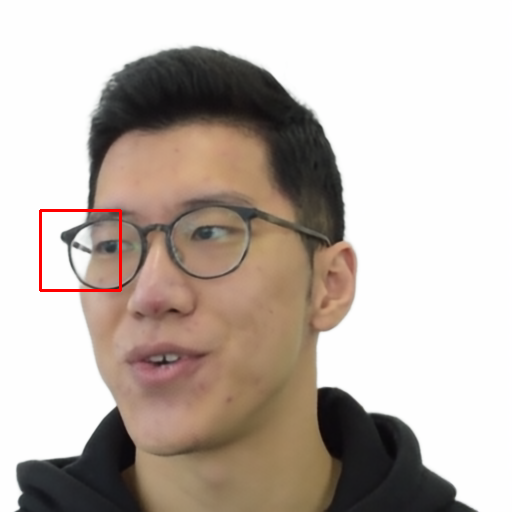}
			\includegraphics[width=\linewidth]{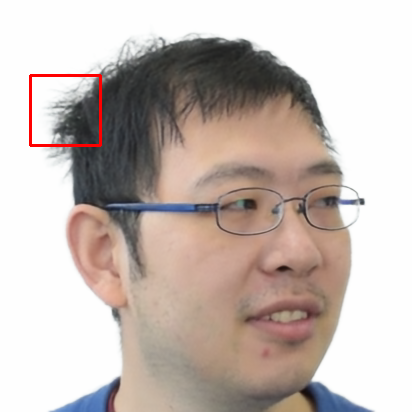}
		\end{minipage}
	}
	%	\subfloat[ours*]{
		%		\begin{minipage}[b]{0.15\linewidth}		
			%			\includegraphics[width=\linewidth]{fig4_42_2-ours0.png}
			%			\includegraphics[width=\linewidth]{fig4_64_3-ours0.png}
			%			\includegraphics[width=\linewidth]{fig4_45_4-ours0.png}
			%			\includegraphics[width=\linewidth]{fig4_0_5-ours0.png}
			%			\includegraphics[width=\linewidth]{fig4_31_6-ours0.png}
			%		\end{minipage}
		%	}
	\subfloat[reference]{
		\begin{minipage}[b]{0.23\linewidth}		
			\includegraphics[width=\linewidth]{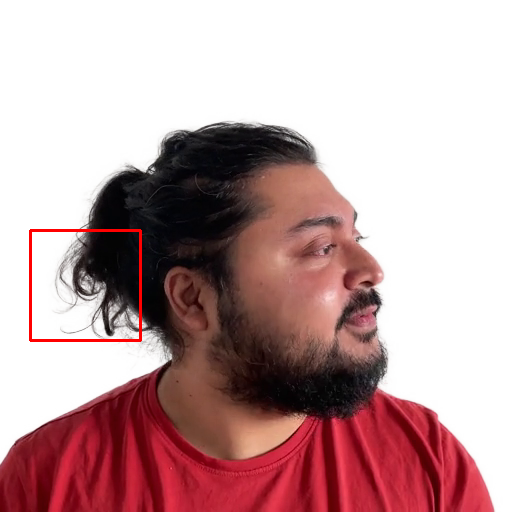}
			\includegraphics[width=\linewidth]{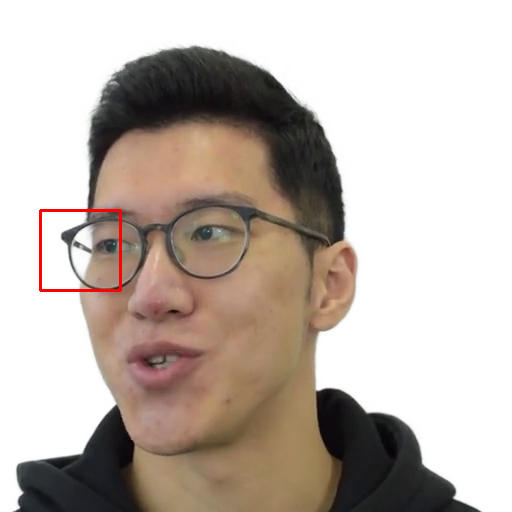}
			\includegraphics[width=\linewidth]{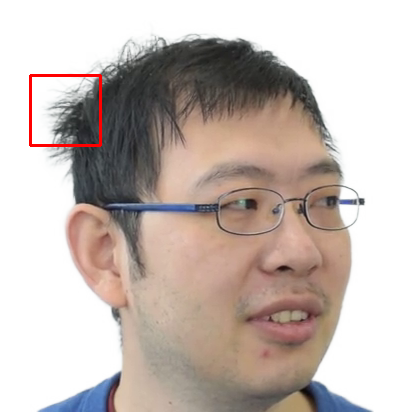}
		\end{minipage}
	}
	\caption{Visualization of each component in PSAvatar. (a) shows the learned point-based shape model. (b) visualizes the 3D Gaussian, which shows improved representation flexibility over PSM. (c) and (d) are the rendered and ground truth image, respectively.}
	\label{fig:four}	
\end{figure}

% added by zy
\begin{table}[]	
	\caption{Quantitative comparison with state-of-the-art methods.
		\colorbox[HTML]{E1EFDB}{Green} and \colorbox[HTML]{FFF2CE}{yellow} indicate the best and the second, respectively.}
	\label{table:one}
	\begin{tabular}{llll}
		\hline \hline
		Methods      & \multicolumn{1}{c}{PSNR $\uparrow$}        & \multicolumn{1}{c}{SSIM $\uparrow $}   & \multicolumn{1}{c}{LPIPS $\downarrow $}    \\  \hline 
		SplattingAvatar \cite{Shao24}                      
		& 21.4032                       
		& 0.8478                                            
		& 0.1863                        \\
		PointAvatar \cite{Zheng23}                   
		& \cellcolor[HTML]{FFF2CE}25.5630                       
		& 0.8920                                            
		& 0.0891                         \\
		BakedAvatar \cite{Duan23}                      
		& 24.9922                       
		& \cellcolor[HTML]{FFF2CE}0.9005                                            
		& \cellcolor[HTML]{FFF2CE}0.0755                         \\
		Ours       
		& \cellcolor[HTML]{E1EFDB}29.9048 
		& \cellcolor[HTML]{E1EFDB}0.9343
		& \cellcolor[HTML]{E1EFDB}0.0497 \\ \hline %\hline
		INSTA \cite{Zielonka23}                        
		& 23.2420                          
		& 0.8886                                         
		& 0.1148                         \\
		FlashAvatar \cite{Xiang2023}                        
		& \cellcolor[HTML]{FFF2CE}26.1108                         
		& \cellcolor[HTML]{FFF2CE}0.9162                                           
		& \cellcolor[HTML]{FFF2CE}0.0790                         \\
		Ours (no cloth)
		& \cellcolor[HTML]{E1EFDB}31.5766 
		& \cellcolor[HTML]{E1EFDB}0.9494 
		& \cellcolor[HTML]{E1EFDB}0.0488 \\ \hline \hline
	\end{tabular}
\end{table}

\subsection{Rendering} \label{Rendering}

To represent the fine detail and model the appearance, PSAvatar employs 3D Gaussian in combination with the PSM.
Specifically, each Gaussian is parameterized by its rotation matrix $R^{'}$, anisotropic scaling matrix $S$, color $c$ and opacity $\sigma$. In contrast to that in PSM, the color $c$ for the Gaussian is modeled by the spherical harmonics.

As shown in Equation (\ref{eqn:07}), samples in the PSM are obtained based on the local coordinate determined by each mesh. To achieve rendering, Gaussians are supposed to be transformed from the local coordinate to the global coordinate by:
\begin{equation}
	\label{eqn:10}
	R = R^{'}R^i
\end{equation}
where $R$ is the rotation matrix of the Gaussian in the global coordinate, and $R^i$ is the local rotation matrix determined by the $i$-th mesh, which is calculated by the barycentric interpolation as well:
\begin{equation}
	\label{eqn:11}
	R^i = \sum_{j=0}^{2}\alpha_j^i R_j^{i}
\end{equation}
where $R_{j}^{i}$ with $j=$\{0, 1, 2\} is the rotation matrix of each vertex that can be derived by $W(\cdot)$ in Equation (\ref{eqn:04}), \textit{i.e.}, $R_{j}^{i}$ corresponds to the rotation part of $W(\cdot)$. The directoinal color is calculated based on the spherical harmonics which is rotated from the local coordinate to the global one using the method in \cite{Nowrouzezahrai12}. %The spherical harmonics conduct rotation as well using the method introduced in \cite{Nowrouzezahrai12}.
Eventually, the tile rasterizer in Equation (\ref{eqn:03}) splats the 3D Gaussians onto the screen to implement the rendering.
Empirically, we have found that the performance of PSAvatar can be further improved with the guidance of the enhancement network.
Considering this, a U-net \cite{Ronneberger15} based enhancement is applied to the rendered image for improving the visual quality.

\subsection{Optimization and Regularization}
The tile rasterizer in Equation (\ref{eqn:03}) will output a rendered image  $I_r$, and the U-net perform  enhancement on the rendered image and produce the ultimate output $I_{enh}$.
The RGB loss constrains the output image in the pixel domain:
\begin{equation}
	\label{eqn:12}
	\mathcal{L}_{RGB}= \left \| I_{r} - I_{GT} \right \| +\left \| I_{enh} - I_{GT} \right \|
\end{equation}
where $I_{GT}$ is the ground truth.
Analogous to prior work \cite{Zheng23}, we adopt a VGG feature loss:
\begin{equation}
	\label{eqn:13}
	\mathcal{L}_{vgg}= \left \| F_{vgg}(I_r) - F_{vgg}(I_{GT}) \right \| +\left \| F_{vgg}(I_{enh}) - F_{vgg}(I_{GT}) \right \|
\end{equation}
where $F_{vgg}$ calculates the features from the first four layers of a pre-trained VGG network \cite{Simonyan14}.
To avoid the scaling vector $s$ growing unbounded,  we regularize the scaling vector $s$ by:
\begin{equation}
	\label{eqn:15}
	\mathcal{L}_{scaling}= \left \|s \right \|
\end{equation}

Formally, the training objective for supervising PSAvatar is defined as:
\begin{equation}
	\label{eqn:16}
	\mathcal{L} = \mathcal{L}_{RGB} 
	+ \lambda_1 \mathcal{L}_{vgg}
	+\lambda_2\mathcal{L}_{scaling}
\end{equation}
where $\lambda_1$ and $\lambda_2$ are taken as 0.1.

%%zy
\begin{figure*}	[!t]
	\centering
	\subfloat[PointAvatar]{
		\begin{minipage}[b]{0.13\linewidth}
			\includegraphics[width=\linewidth]{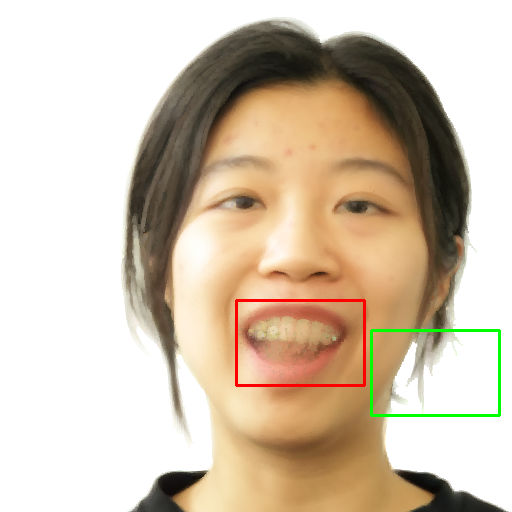}
			\includegraphics[width=0.5\linewidth]{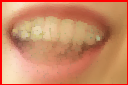}\includegraphics[width=0.5\linewidth]{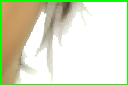}			
			\includegraphics[width=\linewidth]{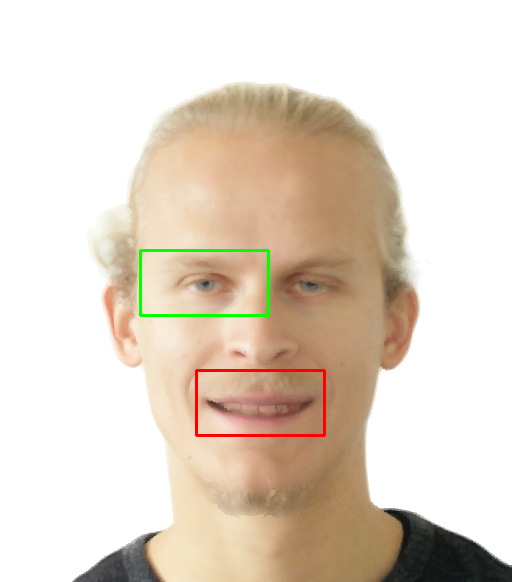}
			\includegraphics[width=0.5\linewidth]{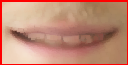}\includegraphics[width=0.5\linewidth]{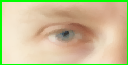}			
			\includegraphics[width=\linewidth]{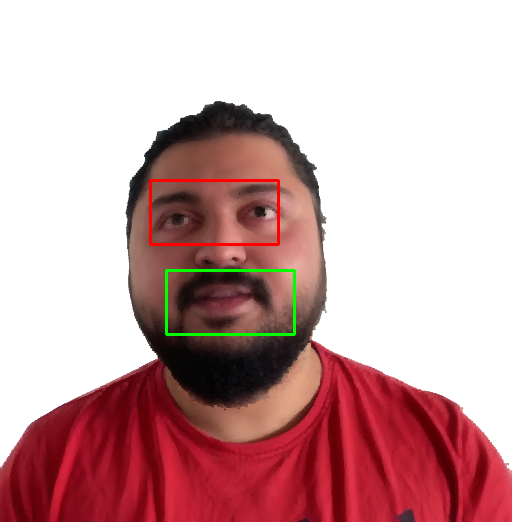}
			\includegraphics[width=0.5\linewidth]{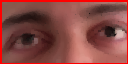}\includegraphics[width=0.5\linewidth]{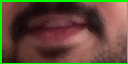}			
			\includegraphics[width=\linewidth]{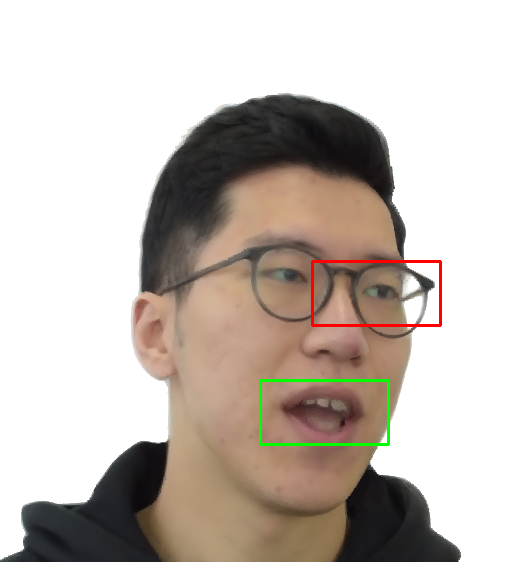}
			\includegraphics[width=0.5\linewidth]{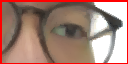}\includegraphics[width=0.5\linewidth]{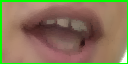}
			\includegraphics[width=\linewidth]{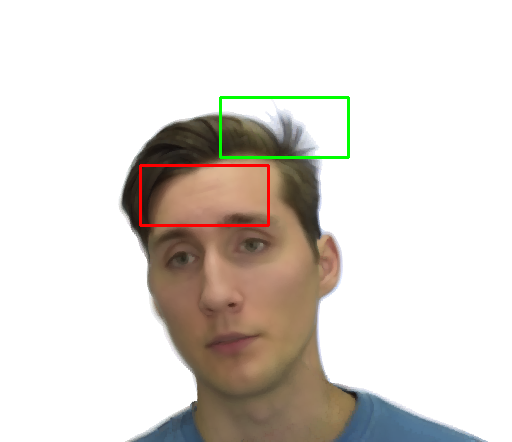}
			\includegraphics[width=0.5\linewidth]{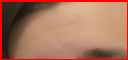}\includegraphics[width=0.5\linewidth]{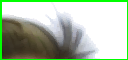}
			\includegraphics[width=\linewidth]{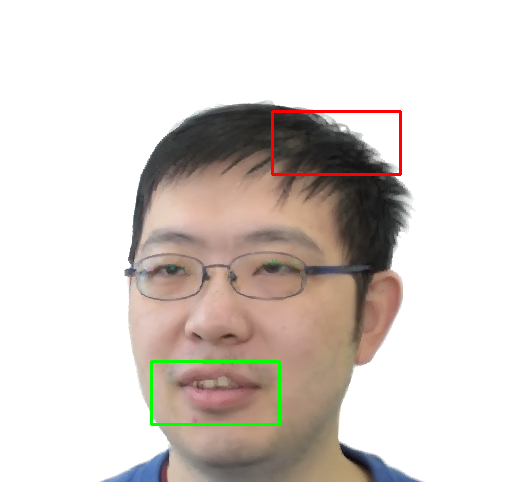}
			\includegraphics[width=0.5\linewidth]{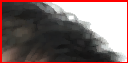}\includegraphics[width=0.5\linewidth]{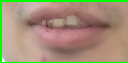}
		\end{minipage}
	}
	\subfloat[INSTA]{
		\begin{minipage}[b]{0.13\linewidth}
			\includegraphics[width=\linewidth]{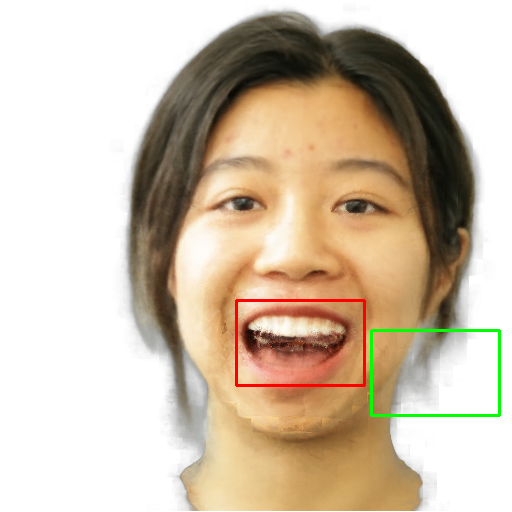}
			\includegraphics[width=0.5\linewidth]{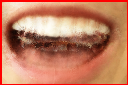}\includegraphics[width=0.5\linewidth]{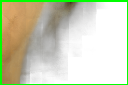}
			\includegraphics[width=\linewidth]{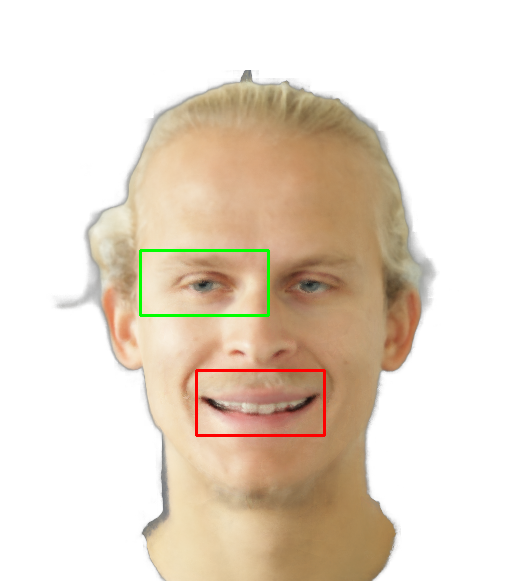}
			\includegraphics[width=0.5\linewidth]{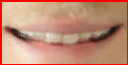}\includegraphics[width=0.5\linewidth]{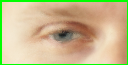}
			\includegraphics[width=\linewidth]{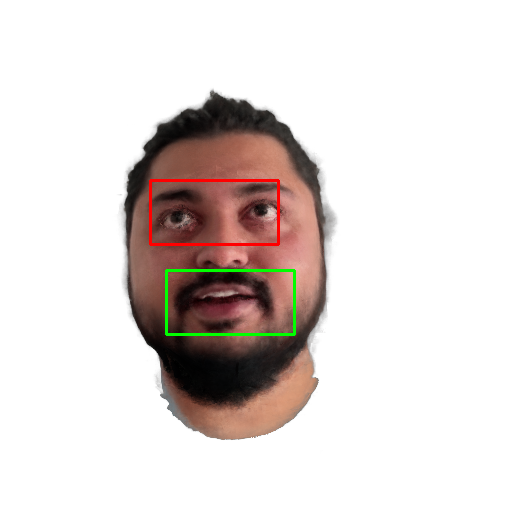}
			\includegraphics[width=0.5\linewidth]{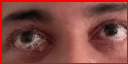}\includegraphics[width=0.5\linewidth]{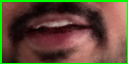}
			\includegraphics[width=\linewidth]{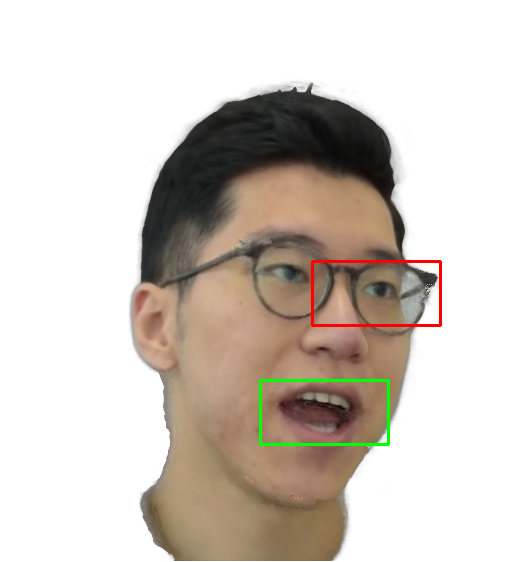}
			\includegraphics[width=0.5\linewidth]{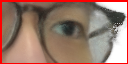}\includegraphics[width=0.5\linewidth]{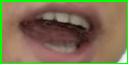}
			\includegraphics[width=\linewidth]{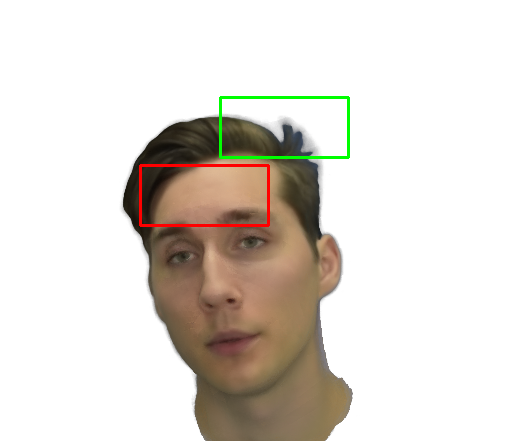}
			\includegraphics[width=0.5\linewidth]{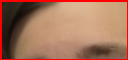}\includegraphics[width=0.5\linewidth]{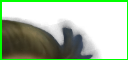}
			\includegraphics[width=\linewidth]{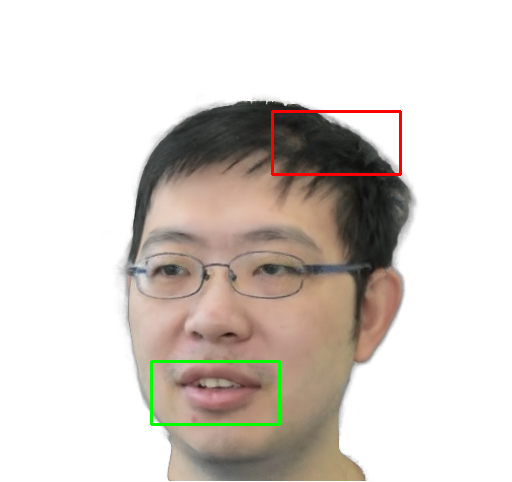}
			\includegraphics[width=0.5\linewidth]{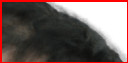}\includegraphics[width=0.5\linewidth]{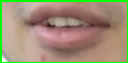}
		\end{minipage}
	}
	\subfloat[BakedAvatar]{
		\begin{minipage}[b]{0.13\linewidth}
			\includegraphics[width=\linewidth]{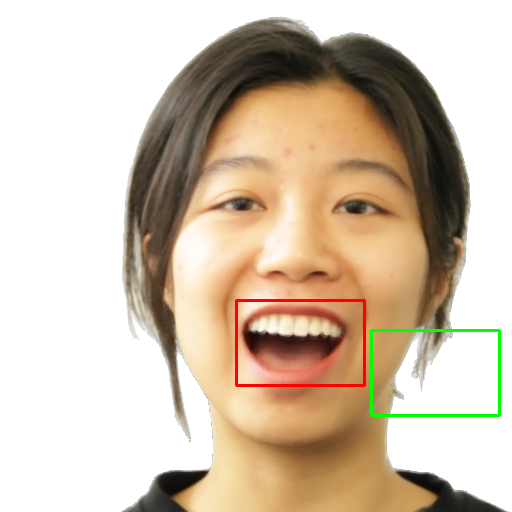}
			\includegraphics[width=0.5\linewidth]{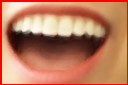}\includegraphics[width=0.5\linewidth]{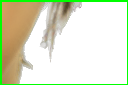}
			\includegraphics[width=\linewidth]{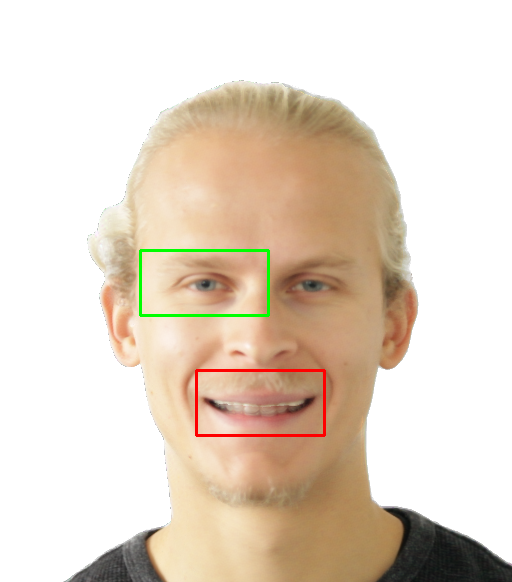}
			\includegraphics[width=0.5\linewidth]{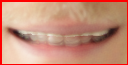}\includegraphics[width=0.5\linewidth]{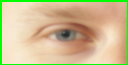}
			\includegraphics[width=\linewidth]{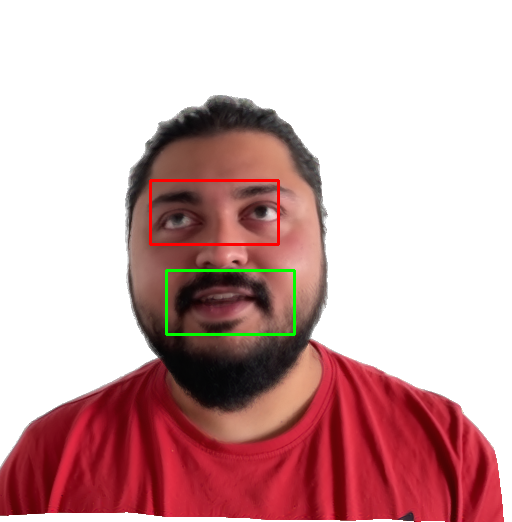}
			\includegraphics[width=0.5\linewidth]{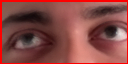}\includegraphics[width=0.5\linewidth]{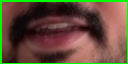}
			\includegraphics[width=\linewidth]{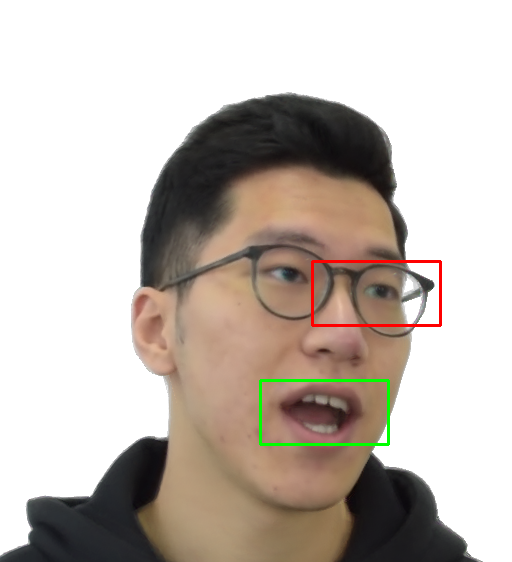}
			\includegraphics[width=0.5\linewidth]{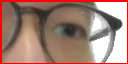}\includegraphics[width=0.5\linewidth]{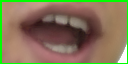}
			\includegraphics[width=\linewidth]{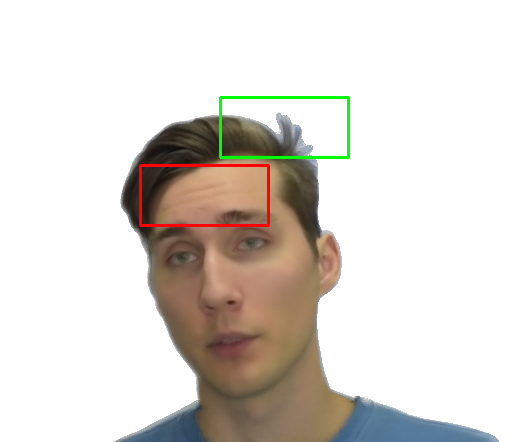}
			\includegraphics[width=0.5\linewidth]{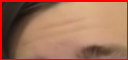}\includegraphics[width=0.5\linewidth]{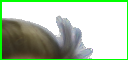}
			\includegraphics[width=\linewidth]{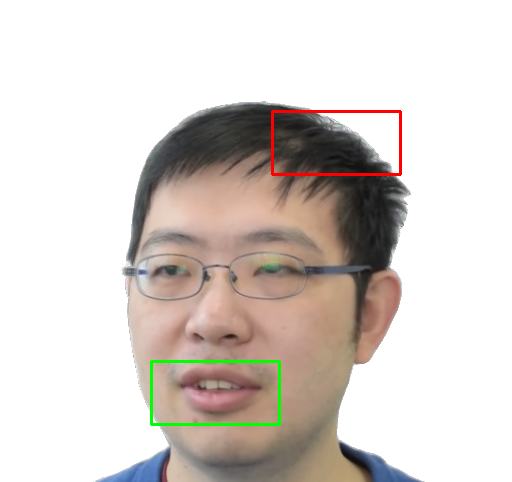}
			\includegraphics[width=0.5\linewidth]{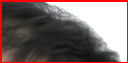}\includegraphics[width=0.5\linewidth]{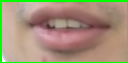}
		\end{minipage}
	}
	\subfloat[SplattingAvatar]{
		\begin{minipage}[b]{0.13\linewidth}
			\includegraphics[width=\linewidth]{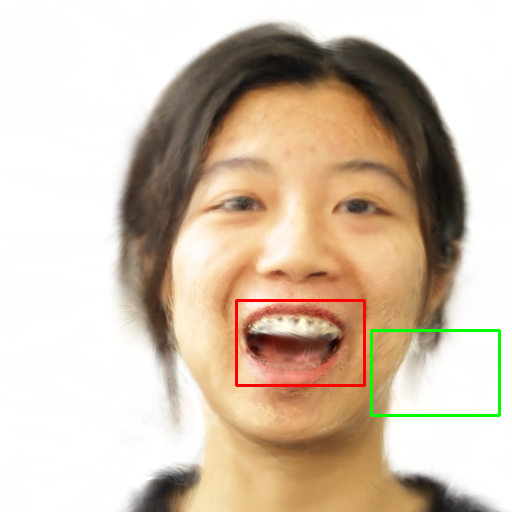}
			\includegraphics[width=0.5\linewidth]{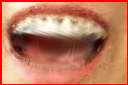}\includegraphics[width=0.5\linewidth]{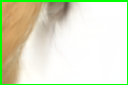}
			\includegraphics[width=\linewidth]{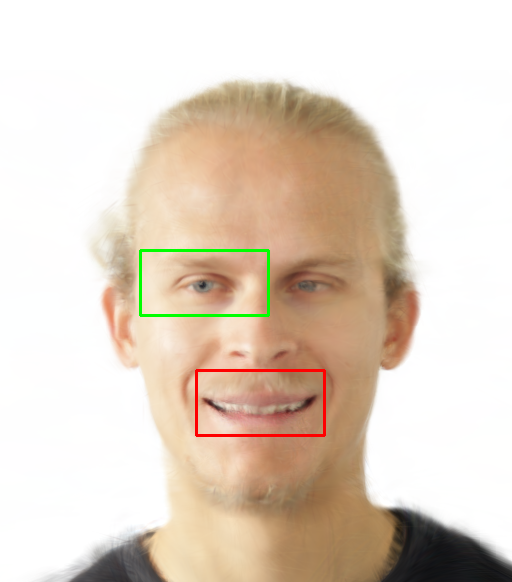}
			\includegraphics[width=0.5\linewidth]{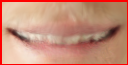}\includegraphics[width=0.5\linewidth]{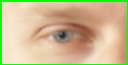}
			\includegraphics[width=\linewidth]{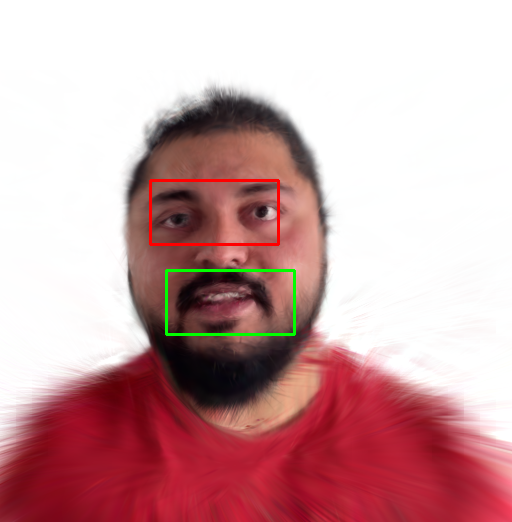}
			\includegraphics[width=0.5\linewidth]{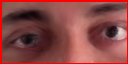}\includegraphics[width=0.5\linewidth]{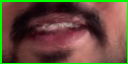}
			\includegraphics[width=\linewidth]{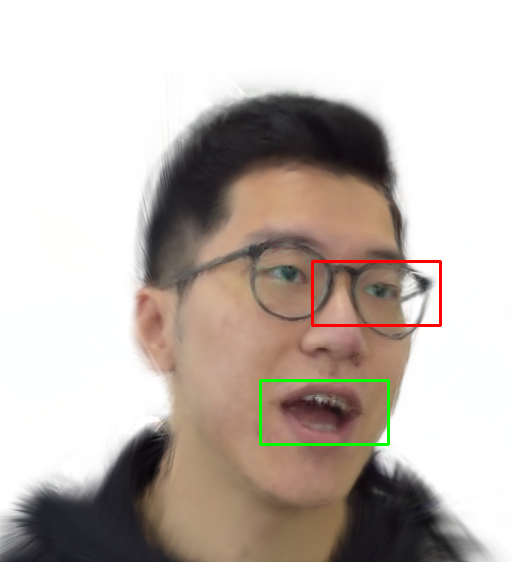}
			\includegraphics[width=0.5\linewidth]{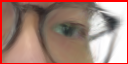}\includegraphics[width=0.5\linewidth]{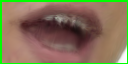}
			\includegraphics[width=\linewidth]{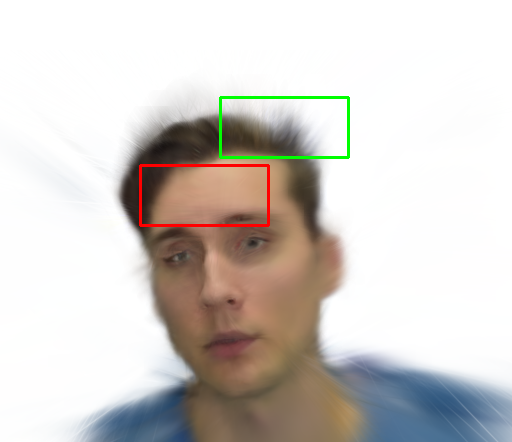}
			\includegraphics[width=0.5\linewidth]{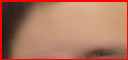}\includegraphics[width=0.5\linewidth]{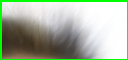}
			\includegraphics[width=\linewidth]{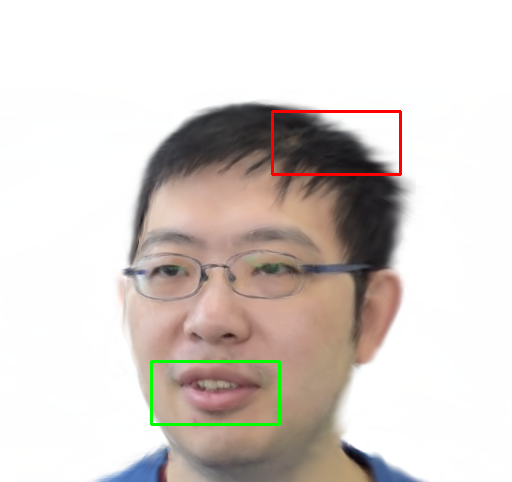}
			\includegraphics[width=0.5\linewidth]{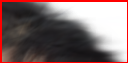}\includegraphics[width=0.5\linewidth]{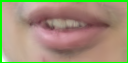}
		\end{minipage}
	}
	\subfloat[FlashAvatar]{
		\begin{minipage}[b]{0.13\linewidth}
			\includegraphics[width=\linewidth]{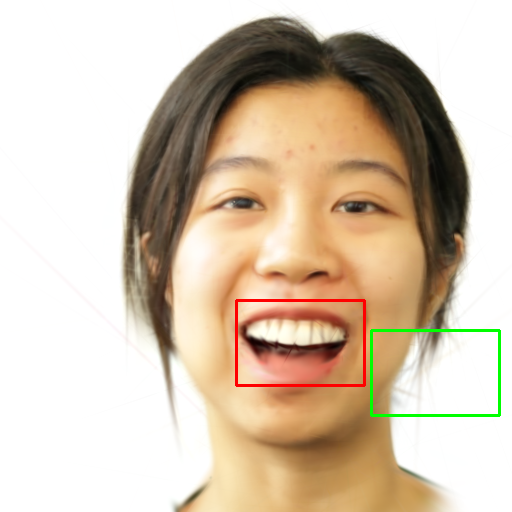}
			\includegraphics[width=0.5\linewidth]{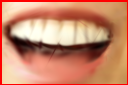}\includegraphics[width=0.5\linewidth]{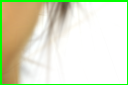}
			\includegraphics[width=\linewidth]{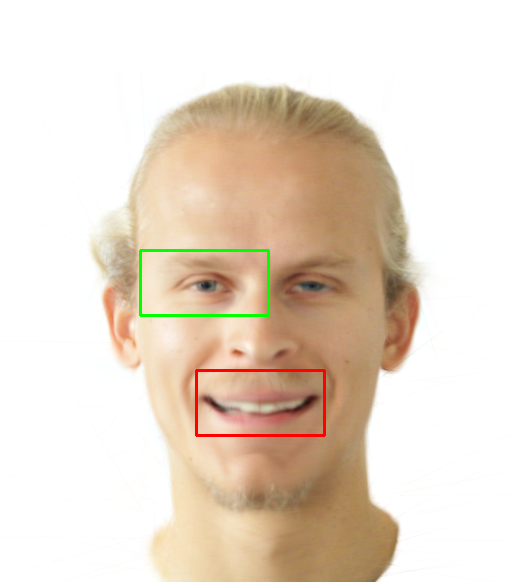}
			\includegraphics[width=0.5\linewidth]{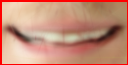}\includegraphics[width=0.5\linewidth]{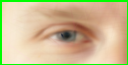}
			\includegraphics[width=\linewidth]{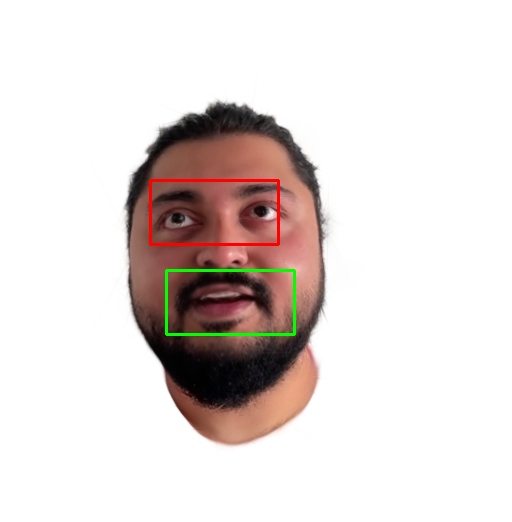}
			\includegraphics[width=0.5\linewidth]{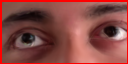}\includegraphics[width=0.5\linewidth]{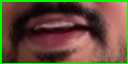}
			\includegraphics[width=\linewidth]{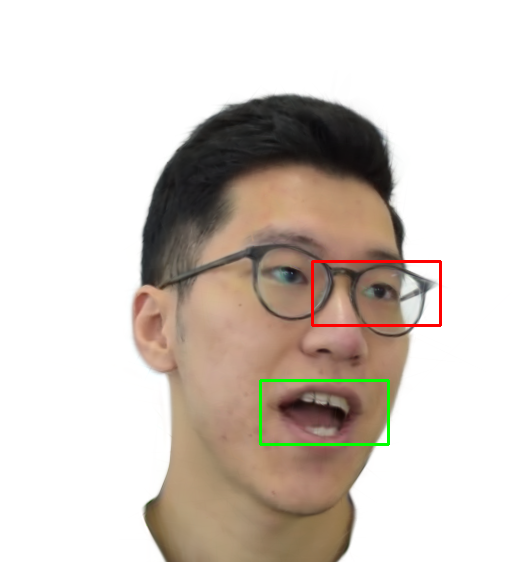}
			\includegraphics[width=0.5\linewidth]{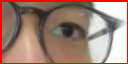}\includegraphics[width=0.5\linewidth]{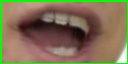}
			\includegraphics[width=\linewidth]{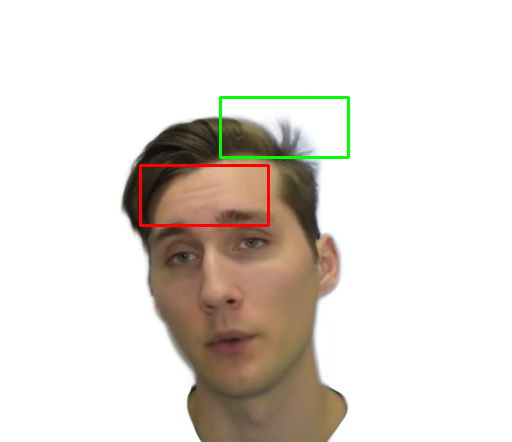}
			\includegraphics[width=0.5\linewidth]{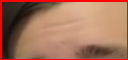}\includegraphics[width=0.5\linewidth]{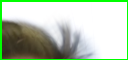}
			\includegraphics[width=\linewidth]{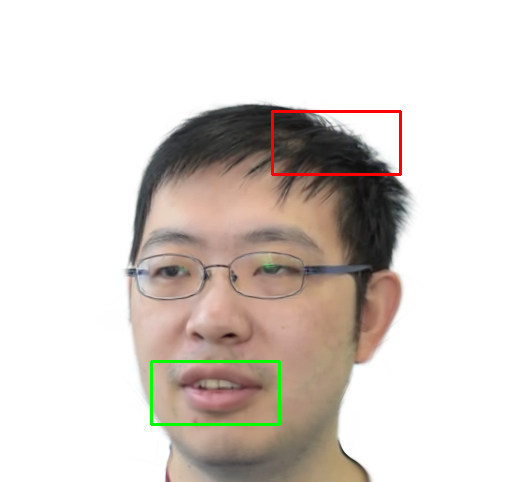}
			\includegraphics[width=0.5\linewidth]{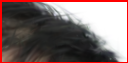}\includegraphics[width=0.5\linewidth]{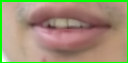}
		\end{minipage}
	}
	\subfloat[Ours]{
		\begin{minipage}[b]{0.13\linewidth}
			\includegraphics[width=\linewidth]{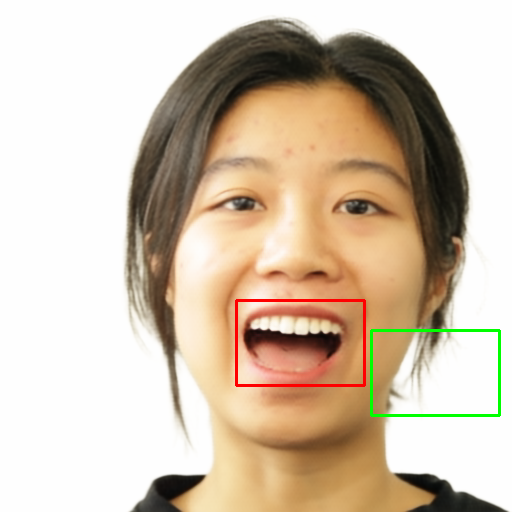}
			\includegraphics[width=0.5\linewidth]{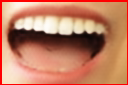}\includegraphics[width=0.5\linewidth]{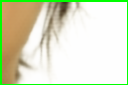}
			\includegraphics[width=\linewidth]{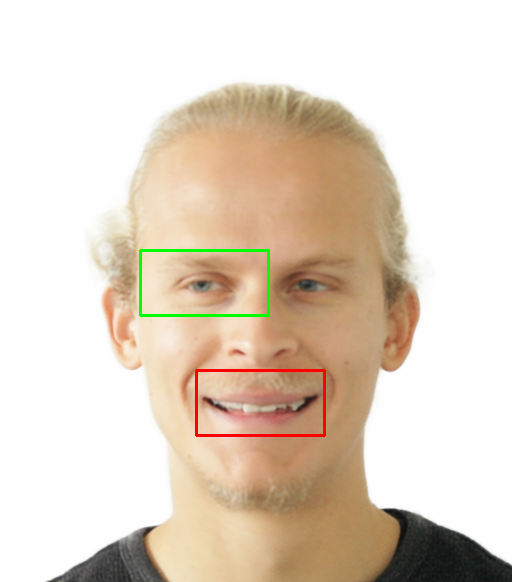}
			\includegraphics[width=0.5\linewidth]{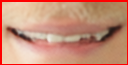}\includegraphics[width=0.5\linewidth]{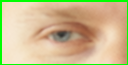}
			\includegraphics[width=\linewidth]{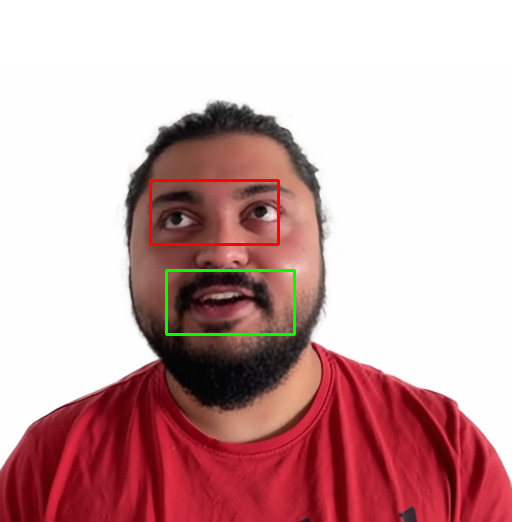}
			\includegraphics[width=0.5\linewidth]{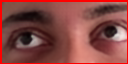}\includegraphics[width=0.5\linewidth]{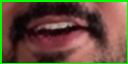}
			\includegraphics[width=\linewidth]{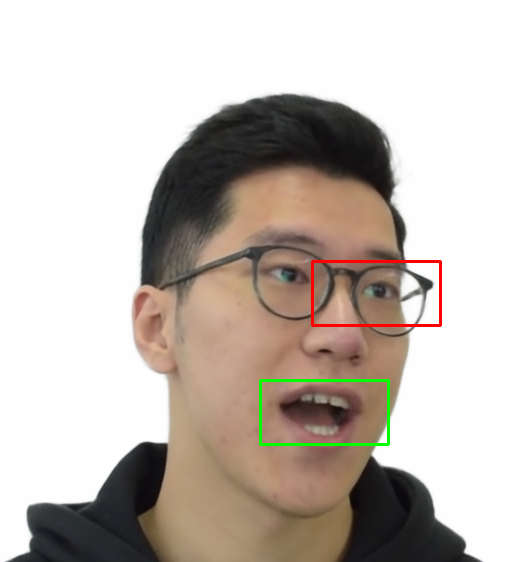}
			\includegraphics[width=0.5\linewidth]{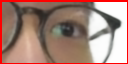}\includegraphics[width=0.5\linewidth]{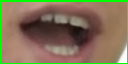}
			\includegraphics[width=\linewidth]{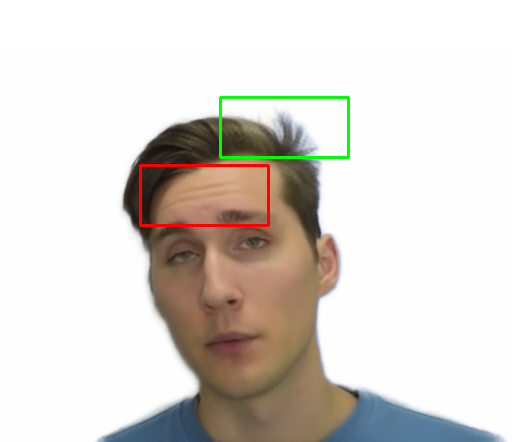}
			\includegraphics[width=0.5\linewidth]{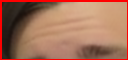}\includegraphics[width=0.5\linewidth]{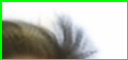}
			\includegraphics[width=\linewidth]{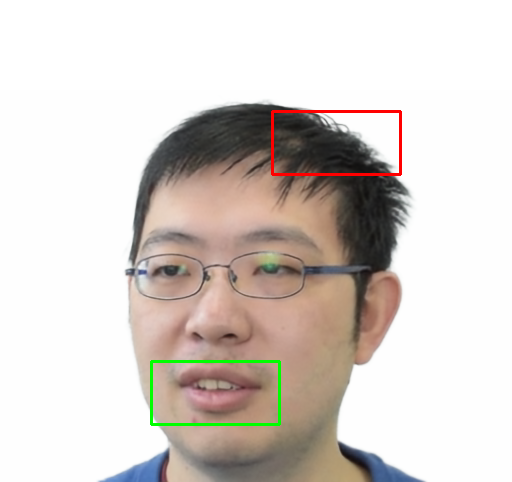}
			\includegraphics[width=0.5\linewidth]{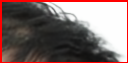}\includegraphics[width=0.5\linewidth]{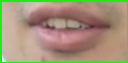}
		\end{minipage}
	}
	\subfloat[GT]{
		\begin{minipage}[b]{0.13\linewidth}
			\includegraphics[width=\linewidth]{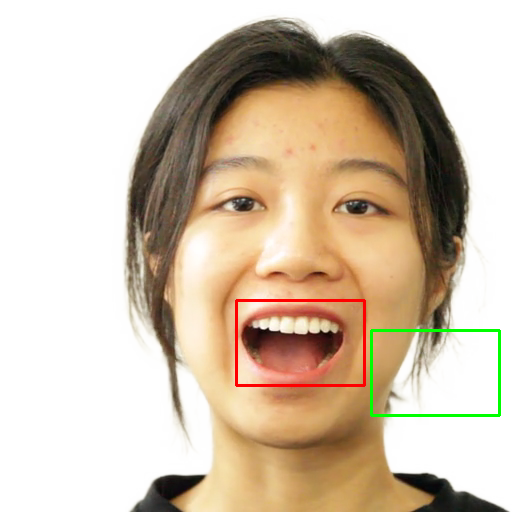}
			\includegraphics[width=0.5\linewidth]{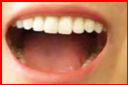}\includegraphics[width=0.5\linewidth]{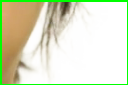}
			\includegraphics[width=\linewidth]{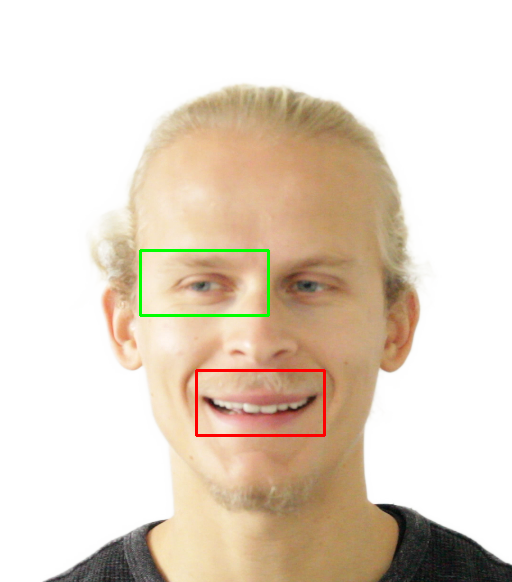}
			\includegraphics[width=0.5\linewidth]{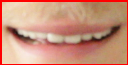}\includegraphics[width=0.5\linewidth]{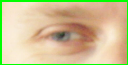}
			\includegraphics[width=\linewidth]{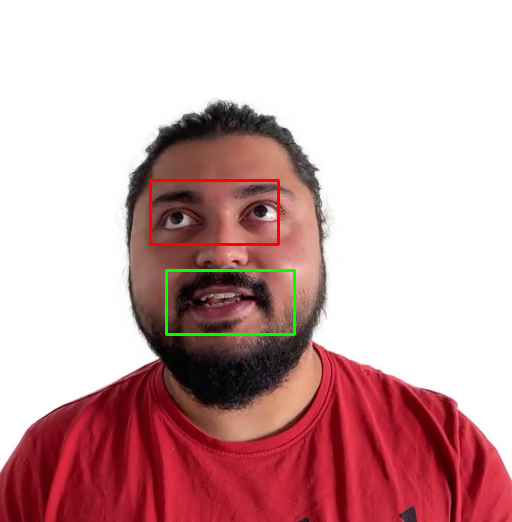}
			\includegraphics[width=0.5\linewidth]{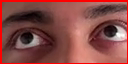}\includegraphics[width=0.5\linewidth]{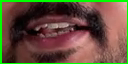}
			\includegraphics[width=\linewidth]{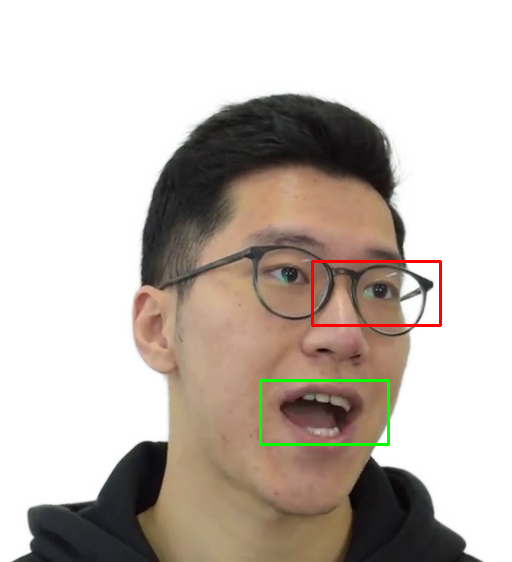}
			\includegraphics[width=0.5\linewidth]{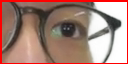}\includegraphics[width=0.5\linewidth]{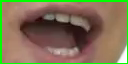}
			\includegraphics[width=\linewidth]{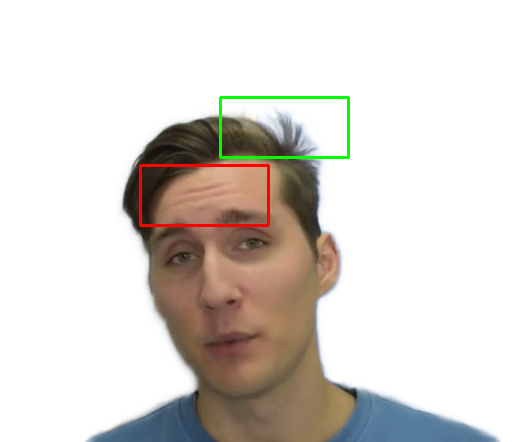}
			\includegraphics[width=0.5\linewidth]{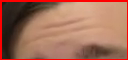}\includegraphics[width=0.5\linewidth]{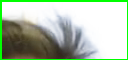}
			\includegraphics[width=\linewidth]{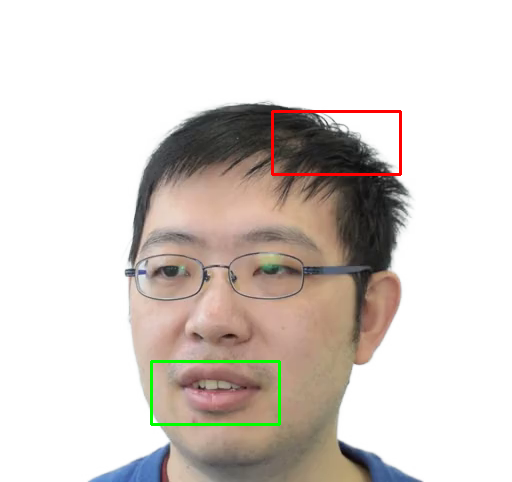}
			\includegraphics[width=0.5\linewidth]{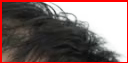}\includegraphics[width=0.5\linewidth]{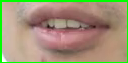}
		\end{minipage}
	}
	\caption{Qualitative comparison on subject 1-6 (from top to bottom). PSAvatar shows improved performances over strong baselines in capturing fine details such as hair strands, teeth, glasses, \textit{etc}.}
	\label{fig:seven}	
\end{figure*}

\section{Experiments}
\subsection{Setup}
PSAvatar is applied for head avatar creation on video recordings of 6 subjects, in which subject 1 (yufeng) and 2 (marcel) are from IMAvatar \cite{Zheng22} (captured by DSLR), subject 3 (soubhik) is from PointAvatar \cite{Zheng23} (captured by smartphone), and subject 4-6 are from NerFace \cite{Gafni21} (captured by DSLR). 
%We follow the protocol of IMAvatar \cite{Zheng22} for the data preprocessing, see more details in the Sup. Mat. 
%The trainset contains 1,000-5,000 frames, while the testset includes frames with novel poses and expressions.
%We follow the protocol of IMAvatar \cite{Zheng22} for the data preprocessing and see more details in the Sup. Mat. 
% we assess our methods by axaming PSNR, SSIM, LPIPS over self-reenactment over diverse ...
%We follow the protocol of IMAvatar \cite{Zheng22} for the data preprocessing and obtain an average of 3,000-4,000 frames for training and 1,000-2,000 frames with novel poses and expressions for testing. 
%For fair comparisons, all these subjects share the same face-tracking algorithm for camera, pose and expression initializations. state-of-the-art 

We compare our method with five representative methods for head avatar creation, each modeling the head avatar in different ways.
INSTA \cite{Zielonka23} establishes point-to-point correspondences between the canonical and deformed space based on the FLAME mesh, and relies on Instant-NGP for radiance field reconstruction.
BakedAvatar \cite{Duan23} learns three implicit fields in canonical space for FLAME-based deformation, and uses differential rasterization to fine-tune the baked textures. PointAvatar  \cite{Zheng23} uses points for 3D represention in the canonical space, while SplattingAvatar \cite{Shao24} learns 3D Gaussians with trainable embedding on the canonical mesh instead. Besides, FlashAvatar \cite{Xiang2023} embeds 3D Gaussians to the mesh based on the fixed UV mapping and learns dynamic offsets to reconstruct a head avatar. 
Note that both INSTA and FlashAvatar use a MICA \cite{Zielonka2022} based preprocessing pipeline and the estimated FLAME parameters are different, to get the best results for those approaches, we adopt the original implementation released by the authors. For all the other methods, we use the same preprocessing pipeline conducted by IMAvatar \cite{Zheng22}.

\subsection{Head Avatar Reconstruction}
Fig. \ref{fig:four} visualizes each component of the introduced PSAvatar.
PSM is capable of capturing the shape variation with poses and expressions, and can reconstruct the volumetric structures, \textit{e.g.,} the glasses are well-reconstructed. 
However, PSM struggles with the modeling of tiny hairs.
To solve this, 3D Gaussians  are employed to enhance the representation flexibility for fine details, and successfully model the complex geometry ignored by PSM.
Generally, PSAvatar can render sharp and realistic images even for extreme poses.
In addition, the supplementary video sequences demonstrate that the reconstructed head avatars show good performances on high fidelity rendering and can be animated in real-time by the pose, expression and  camera parameters.

For quantitative comparison, Table \ref{table:one} lists the conventional metrics measured on the introduced PSAvatar and SOTA baselines.
It is clearly seen that our approach outperforms others by a significant margin in terms of PSNR, SSIM and LPIPS \cite{Zhang18}.
Besides, we evaluate the reconstruction quality via qualitative comparison. As shown in Fig. \ref{fig:seven}, 
PointAvatar generates smooth volumetric effects but struggles with rendering sharp images, \textit{e.g.}, the reconstructed teeth is blurry, and PointAvatar is incapable of reproducing tiny hair found in subject 1, 2 and 6. INSTA deforms points according to the nearest triangle, causing the misalignment between the FLAME mesh and the target geometry which greatly decreases the rendering quality. Hence, the mouth, hair and eye regions synthesized by INSTA suffer from noises. In a different way of head representation, BakedAvatar, based on the deformable multi-layer meshes which could be baked by the learned neutral fields, is theoretically capable of modeling intricate details, but still insufficient to reconstruct the complex hairstyles and the fine-grained teeth. 
For 3D Gaussian-based methods, SplattingAvatar employs a movable binding strategy between 3D Gaussian and head mesh through the trainable embeddings, which potentially introduces geometric inconsistency, thus affecting the performance of 3D representation and the rendering quality. Besides, FlashAvatar performs well on producing geometry-consistent results with the UV mapping, but the representation ability for non-facial structures such as eyeglasses and long thin hair is still restricted by the limited capacity of the offset network.

Owing to the representation capability provided by the point-based shape model, PSAvatar can successfully model  the surface-like topology in the face region and capture complex geometries introduced by diverse hairstyles and accessories.
In addition, PSAvatar gains improved representation capability with the employment of 3D Gaussians during rendering.
As seen in Fig. \ref{fig:seven}, PSAvatar shows plausible results on reproducing intricate details, \textit{e.g.}, the tiny single hair of subject 1, the hair bun of subject 2, the glasses of subject 4 and 6.
Remarkably, PSAvatar can reproduce the expression-dependent wrinkle in the forehead of subject 5, which is much more realistic than the other head avatars and further demonstrates the improved performance of PSAvatar over the existing models.

%%%
\subsection{Cross-identity Reenactment}
\begin{figure*}	[!t]
	\centering
	\subfloat[Source]{
		\begin{minipage}[b]{0.13\linewidth}
			\includegraphics[width=\linewidth]{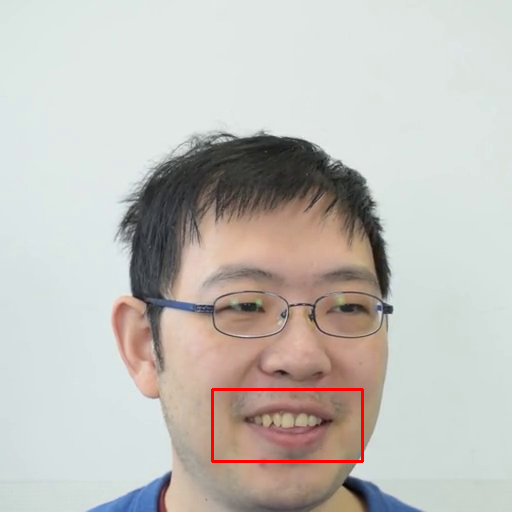}
			\includegraphics[width=0.5\linewidth]{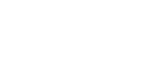}\includegraphics[width=0.5\linewidth]{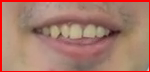}			
			\includegraphics[width=\linewidth]{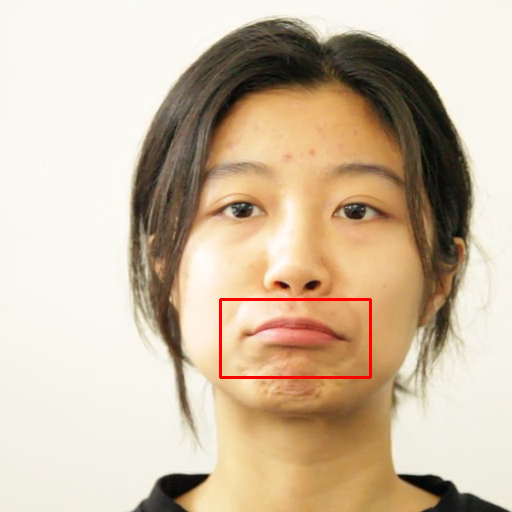}
			\includegraphics[width=0.5\linewidth]{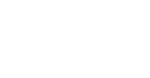}\includegraphics[width=0.5\linewidth]{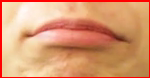}			
		\end{minipage}
	}
	\subfloat[Ours]{
		\begin{minipage}[b]{0.13\linewidth}
			\includegraphics[width=\linewidth]{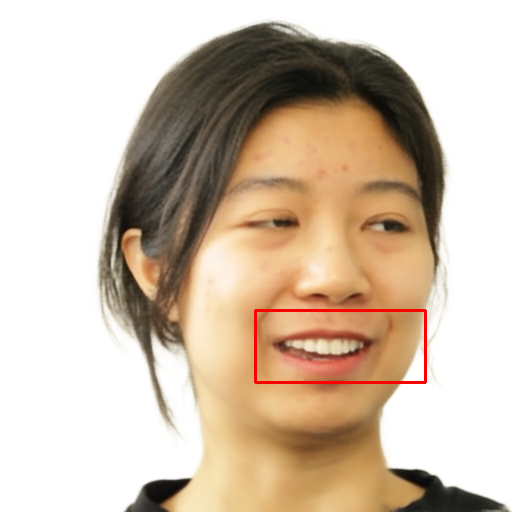}
			\includegraphics[width=0.5\linewidth]{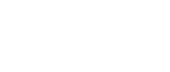}\includegraphics[width=0.5\linewidth]{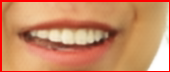}
			\includegraphics[width=\linewidth]{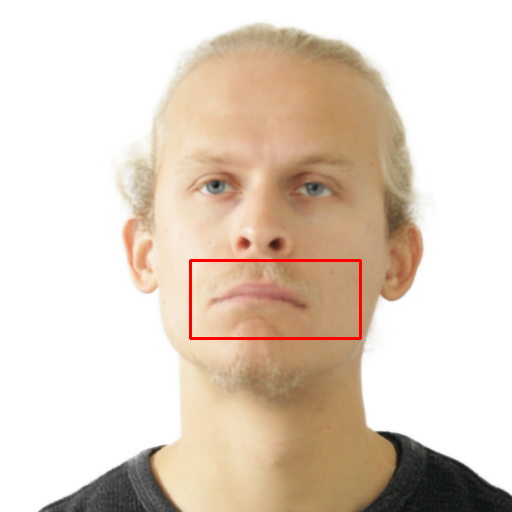}
			\includegraphics[width=0.5\linewidth]{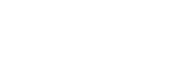}\includegraphics[width=0.5\linewidth]{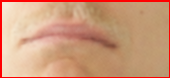}
		\end{minipage}
	}
	\subfloat[INSTA]{
		\begin{minipage}[b]{0.13\linewidth}
			\includegraphics[width=\linewidth]{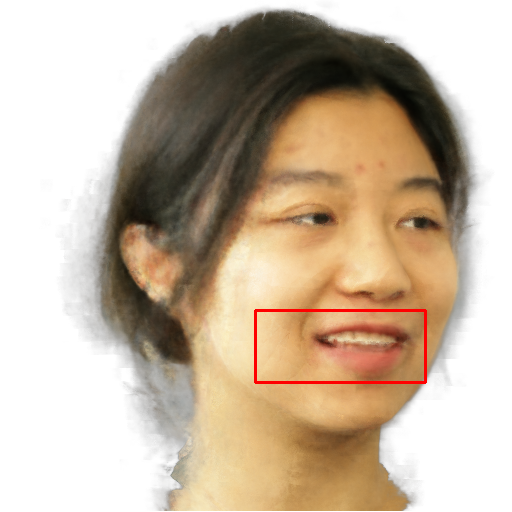}
			\includegraphics[width=0.5\linewidth]{1_driven_w.png}\includegraphics[width=0.5\linewidth]{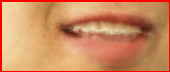}
			\includegraphics[width=\linewidth]{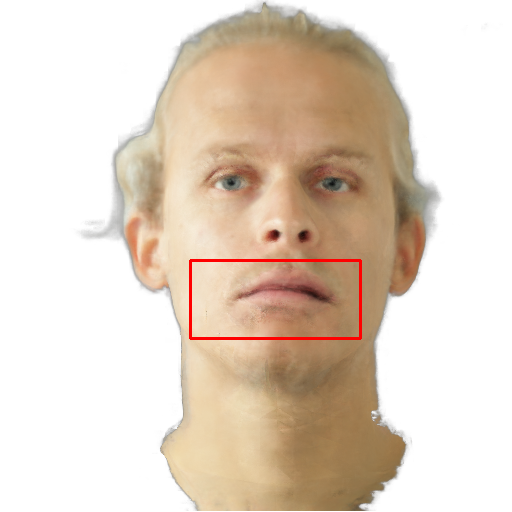}
			\includegraphics[width=0.5\linewidth]{2_driven_w.png}\includegraphics[width=0.5\linewidth]{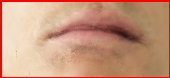}
		\end{minipage}
	}
	\subfloat[BakedAvatar]{
		\begin{minipage}[b]{0.13\linewidth}
			\includegraphics[width=\linewidth]{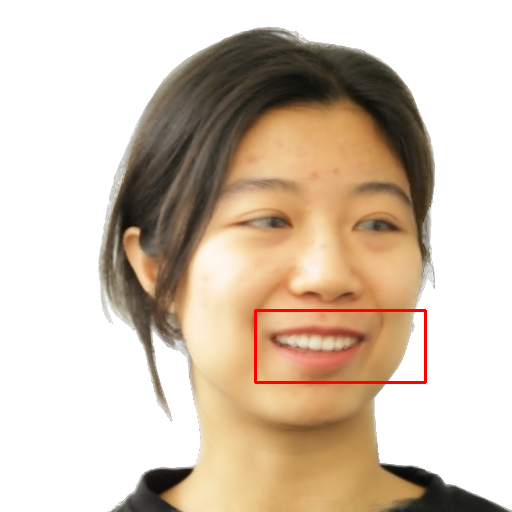}
			\includegraphics[width=0.5\linewidth]{1_driven_w.png}\includegraphics[width=0.5\linewidth]{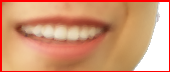}
			\includegraphics[width=\linewidth]{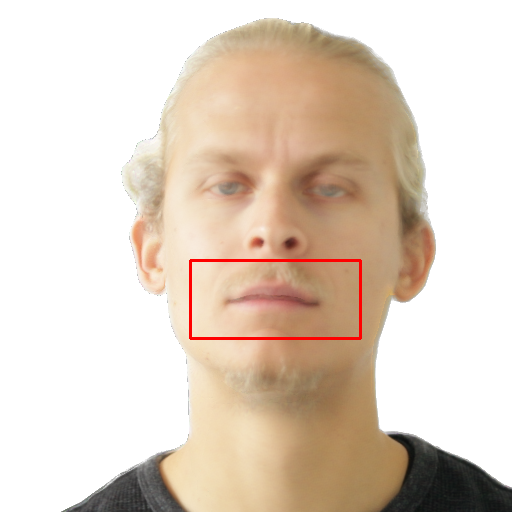}
			\includegraphics[width=0.5\linewidth]{2_driven_w.png}\includegraphics[width=0.5\linewidth]{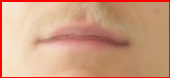}
		\end{minipage}
	}
	\subfloat[SplattingAvatar]{
		\begin{minipage}[b]{0.13\linewidth}
			\includegraphics[width=\linewidth]{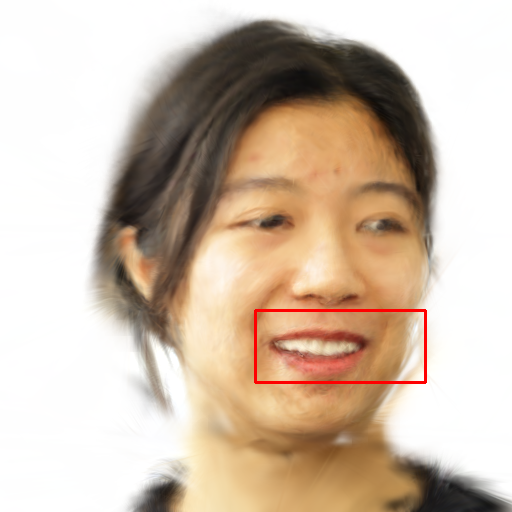}
			\includegraphics[width=0.5\linewidth]{1_driven_w.png}\includegraphics[width=0.5\linewidth]{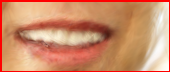}
			\includegraphics[width=\linewidth]{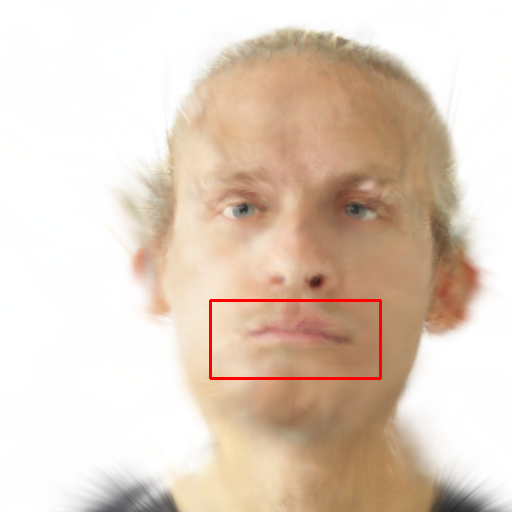}
			\includegraphics[width=0.5\linewidth]{2_driven_w.png}\includegraphics[width=0.5\linewidth]{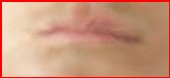}
		\end{minipage}
	}
	\subfloat[PointAvatar]{
		\begin{minipage}[b]{0.13\linewidth}
			\includegraphics[width=\linewidth]{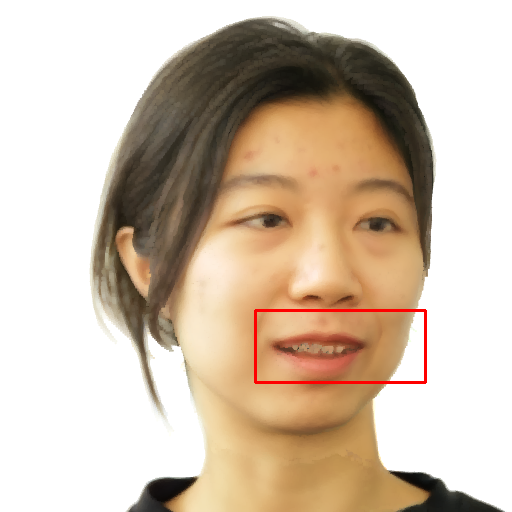}
			\includegraphics[width=0.5\linewidth]{1_driven_w.png}\includegraphics[width=0.5\linewidth]{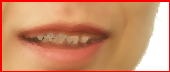}
			\includegraphics[width=\linewidth]{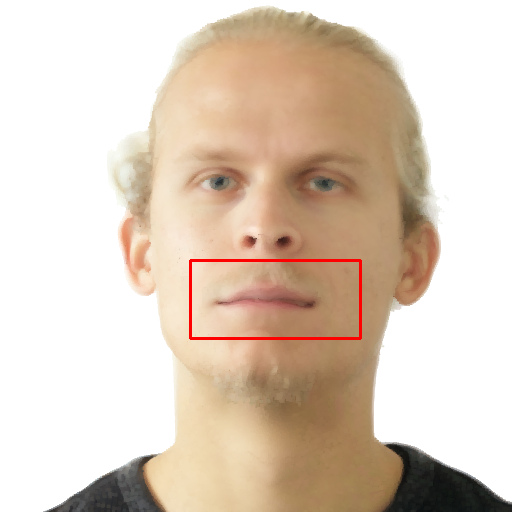}
			\includegraphics[width=0.5\linewidth]{2_driven_w.png}\includegraphics[width=0.5\linewidth]{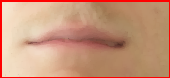}
		\end{minipage}
	}
	\subfloat[FlashAvatar]{
		\begin{minipage}[b]{0.13\linewidth}
			\includegraphics[width=\linewidth]{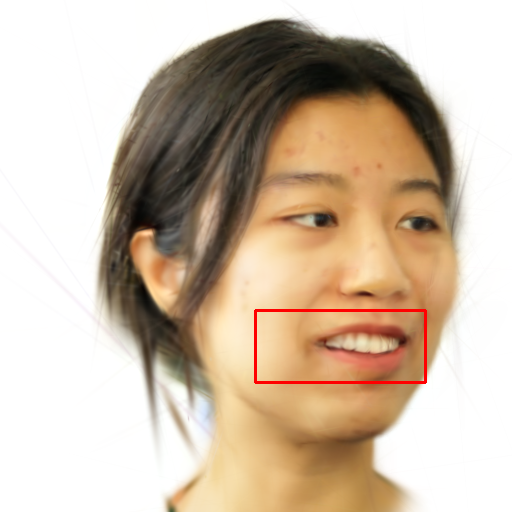}
			\includegraphics[width=0.5\linewidth]{1_driven_w.png}\includegraphics[width=0.5\linewidth]{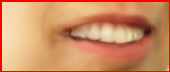}
			\includegraphics[width=\linewidth]{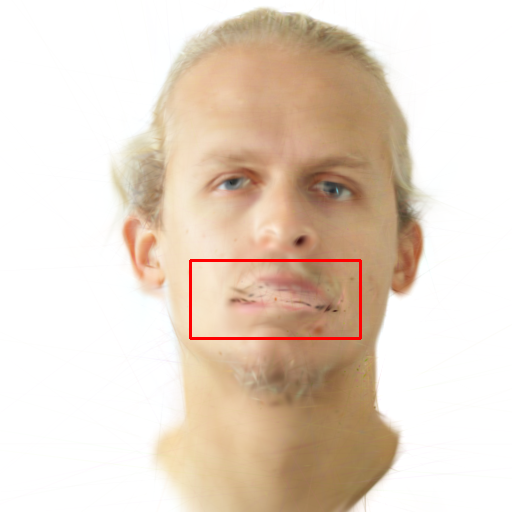}
			\includegraphics[width=0.5\linewidth]{2_driven_w.png}\includegraphics[width=0.5\linewidth]{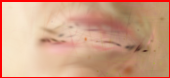}
		\end{minipage}
	}
	\caption{Cross-identity reenactment results from different sources. We use the pose and expression coefficients extracted from the source sequence to animate these target identities. It's clear that PSAvatar outperforms baselines in replicating the corresponding shape variations.}
	\label{fig:renect}	
\end{figure*}

Replication of shape variations caused by different poses and expressions is crucial for head avatar evaluation. Benefit from the efficient combination of our PSM and 3D Gaussian representation, PSAvatar is capable of producing a motion-consistant cross-identity reenactment as well as perserving high-fidelity personalized facial details. The reenactment results of PSAvatar and other representative methods are illustrated as Fig. \ref{fig:renect}. 
%Additionally, thanks to the deformability inherited from FLAME, PSAvatar can be well edited with a wide range of expressions and poses, as well as shapes, \textit{etc.} See the video demonstrations in the Supplementary Materials. 

%% added by zy

\subsection{Ablation Study}\label{Ablation}
%PSAvatar utilizes points to model the pose- and expression-dependent shape variation, and employs Gaussian splats for capturing fine details and efficient rendering. In addition, the U-net based enhancement is applied to improve the visual quality.
%To validate the effectiveness of our method, we conduct experiments to measure the effect of each component, and report the quantitative results in Table \ref{table:two}.
To validate the effectiveness of our method, we conduct experiments to measure the effect of each component.

\begin{figure}[!t]
	\centering
	\subfloat[with $\mathcal{G}$]{
		\begin{minipage}[b]{0.30\linewidth}		
			\includegraphics[width=1\linewidth]{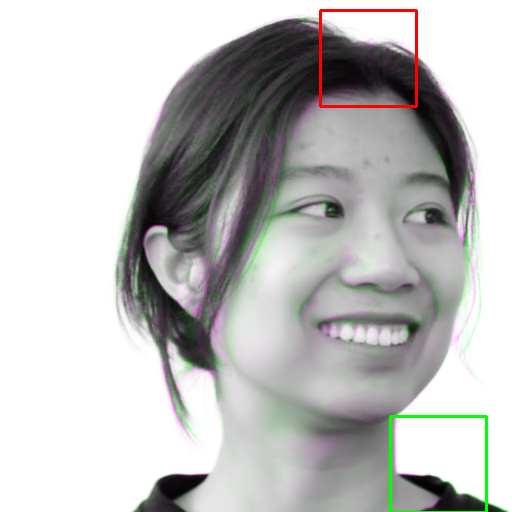}	
		\end{minipage}
		\begin{minipage}[b]{0.15\linewidth}		
			\includegraphics[width=1\linewidth]{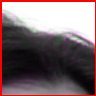}	
			\includegraphics[width=1\linewidth]{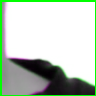}
		\end{minipage}
	}	
	\subfloat[without $\mathcal{G}$]{
		\begin{minipage}[b]{0.30\linewidth}		
			\includegraphics[width=1\linewidth]{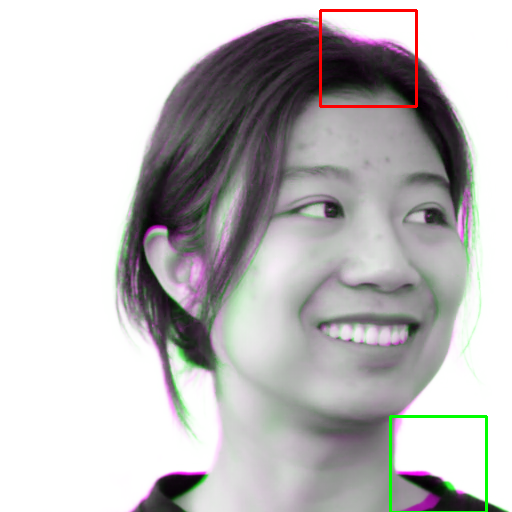}	
		\end{minipage}
		\begin{minipage}[b]{0.15\linewidth}		
			\includegraphics[width=1\linewidth]{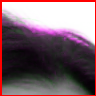}	
			\includegraphics[width=1\linewidth]{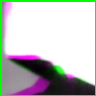}
		\end{minipage}
	}	
	\caption{The effect of $\mathcal{G}$. The highlight pixels indicate the misalignment between the rendered image and the reference. }	
	\label{fig:five}	
	\vspace{-5pt}
\end{figure}

\noindent{\textbf{{$\mathcal{G}$ in Equation (\ref{eqn:05})}}} INSTA \cite{Zielonka23} reconstructs the radiance field of human head based on the FLAME mesh as well, but neglects the misalignment between the FLAME mesh and the target geometry, thus causing the noises in the rendered images (see Fig. \ref{fig:seven}). 
To address this, $\mathcal{G}$ is applied to learn the per-vertex geometry correction.
As seen in Table \ref{table:two} and Fig. \ref{fig:five},
applying $\mathcal{G}$ alleviates the misalignment problem, and contributes to the rendering quality.

\begin{figure}[]
	\centering
	\subfloat[]{
		\begin{minipage}[b]{0.3\linewidth}		
			\includegraphics[width=1\linewidth]{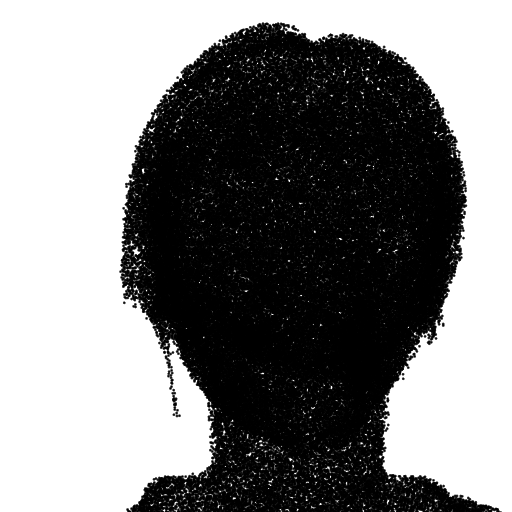}		
		\end{minipage}
	}
	\subfloat[]{
		\begin{minipage}[b]{0.3\linewidth}
			\includegraphics[width=1\linewidth]{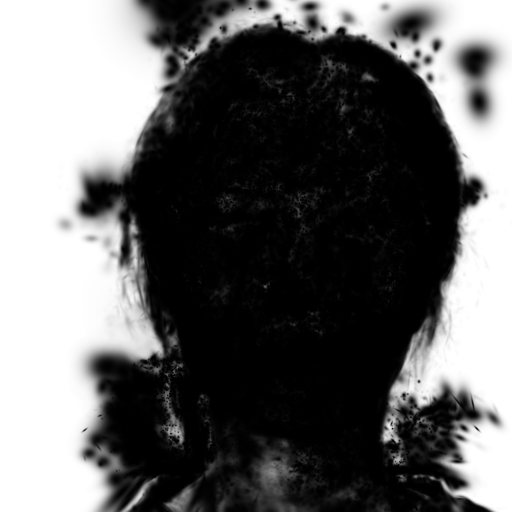}		
		\end{minipage}
	}
	\subfloat[]{
		\begin{minipage}[b]{0.3\linewidth}
			\includegraphics[width=1\linewidth]{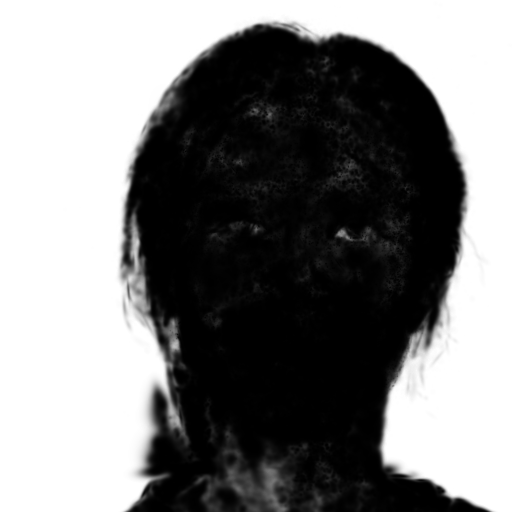}			
		\end{minipage}
	}	
	\caption{The difference between point and 3D Gaussian on modeling shape. (a) shows the points trained with 2 epochs, (b) and (c) show the Gaussians trained with 2 and 10 epochs, respectively. It is clear that points converge quickly to capture the head shape.}	
	\label{fig:six}	
\end{figure}

\noindent{\textbf{{PSM}}}
In order to create the point-based shape model, points are utilized to model the pose- and expression-dependent shape variation. 
Fig. \ref{fig:six} shows the differences between points and 3D Gaussians on the implementation of PSM.
It is clearly seen that points trained with 2 epochs can successfully reconstruct the geometry including the hair strands. However, Gaussian splatting fails to model the shape, even trained with 10 epochs.
The points only optimize the color $c$ and the opacity $\sigma$, in contrast to 3D Gaussians which are additionally parameterized by the rotation and scaling. Hence, points can converge quickly to approximate the shape of head. Considering this, the PSM is achieved based on points instead of Gaussians.

\begin{figure*}[!t]
	\begin{minipage}{0.18\linewidth}
		\centering
		\subfloat[shape]{
			\begin{minipage}[m]{1\linewidth}
				\includegraphics[width=\linewidth]{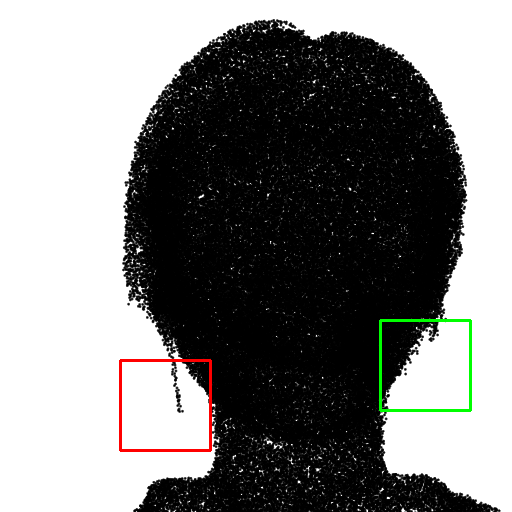}
				\includegraphics[width=0.5\linewidth]{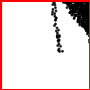}\includegraphics[width=0.5\linewidth]{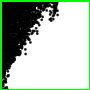}		
			\end{minipage}
		}
	\end{minipage}	
	\medskip
	\begin{minipage}{0.18\linewidth}
		\subfloat[3D Gaussians]{
			\begin{minipage}[b]{1\linewidth}
				\includegraphics[width=\linewidth]{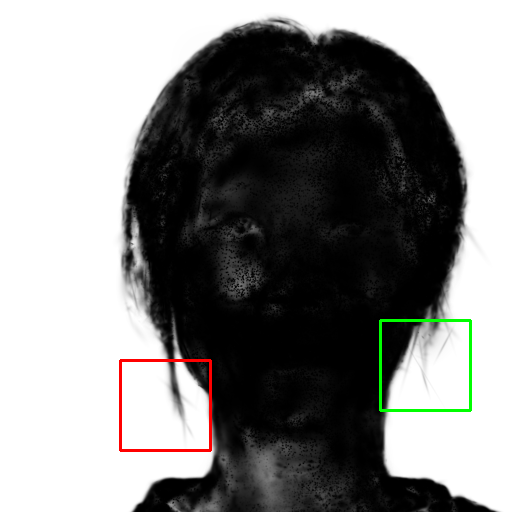}
				\includegraphics[width=0.5\linewidth]{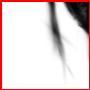}\includegraphics[width=0.5\linewidth]{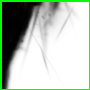}		
			\end{minipage}
		}	
		
		\subfloat[points]{
			\begin{minipage}[b]{1\linewidth}
				\includegraphics[width=\linewidth]{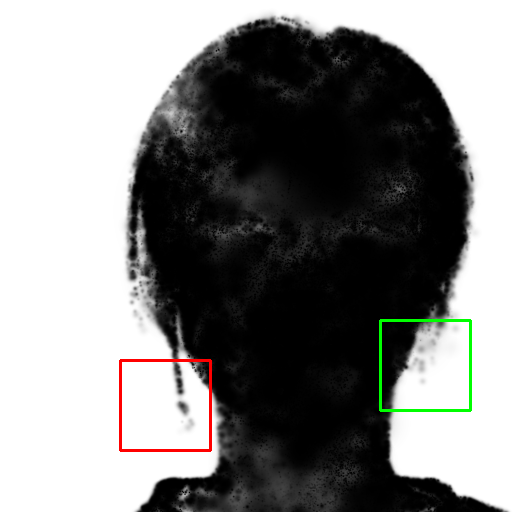}
				\includegraphics[width=0.5\linewidth]{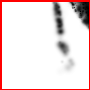}\includegraphics[width=0.5\linewidth]{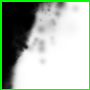}		
			\end{minipage}
		}
	\end{minipage}	
	\medskip
	\begin{minipage}{0.18\linewidth}
		\subfloat[GS]{
			\begin{minipage}[b]{1\linewidth}
				\includegraphics[width=\linewidth]{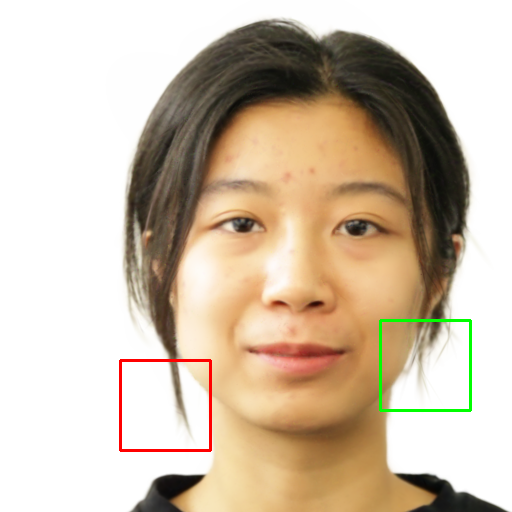}
				\includegraphics[width=0.5\linewidth]{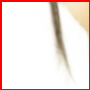}\includegraphics[width=0.5\linewidth]{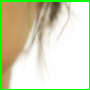}		
			\end{minipage}
		}	
		
		\subfloat[PS]{
			\begin{minipage}[b]{1\linewidth}
				\includegraphics[width=\linewidth]{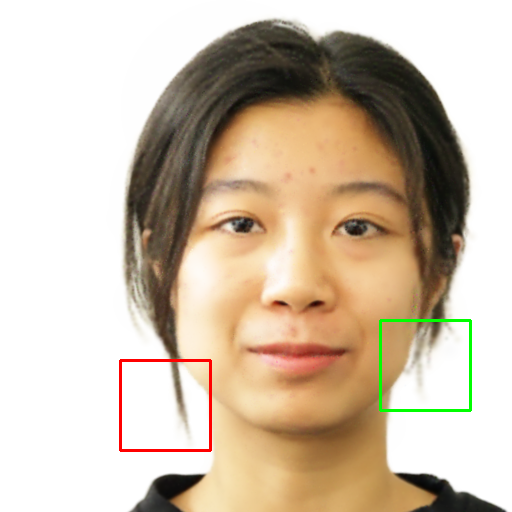}
				\includegraphics[width=0.5\linewidth]{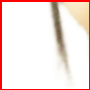}\includegraphics[width=0.5\linewidth]{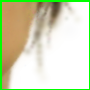}		
			\end{minipage}
		}
	\end{minipage}	
	\medskip
	\begin{minipage}{0.18\linewidth}
		\subfloat[GS+ENH]{
			\begin{minipage}[b]{1\linewidth}
				\includegraphics[width=\linewidth]{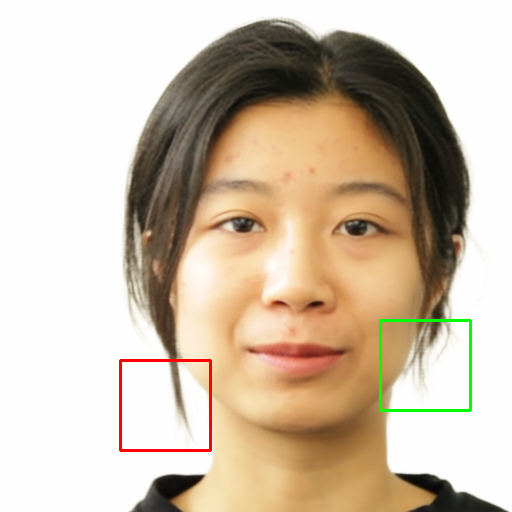}
				\includegraphics[width=0.5\linewidth]{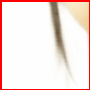}\includegraphics[width=0.5\linewidth]{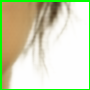}		
			\end{minipage}
		}	
		
		\subfloat[PS+ENH]{
			\begin{minipage}[b]{1\linewidth}
				\includegraphics[width=\linewidth]{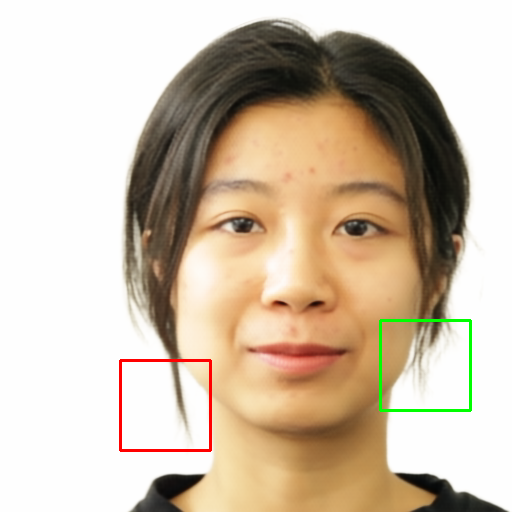}
				\includegraphics[width=0.5\linewidth]{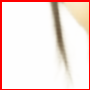}\includegraphics[width=0.5\linewidth]{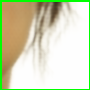}		
			\end{minipage}
		}
	\end{minipage}	
	\medskip
	\begin{minipage}{0.18\linewidth}
		\centering
		\subfloat[ground truth]{
			\begin{minipage}[m]{1\linewidth}
				\includegraphics[width=\linewidth]{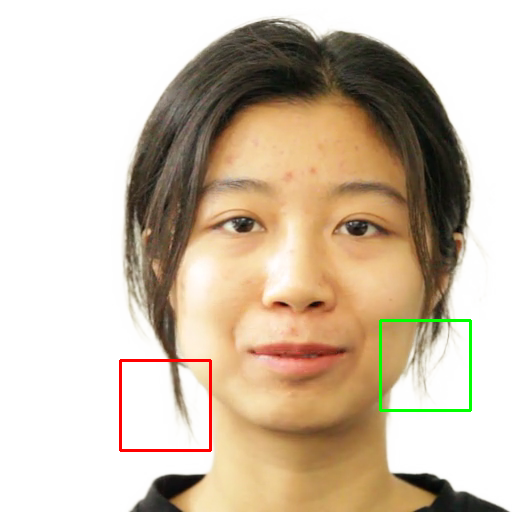}
				\includegraphics[width=0.5\linewidth]{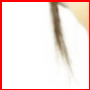}\includegraphics[width=0.5\linewidth]{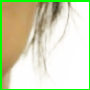}		
			\end{minipage}
		}
	\end{minipage}		
	
	\caption{The effect of each component. (a) shows the shape obtained by the PSM, (b), (d) and (f) are obtained by Gaussian splatting, where 'GS' refers to the image rendered by Gaussian splatting, 'ENH' refers to the enhancement operation, while (c), (e) and (g) are based on point splatting, where 'PS' refers to the image rendered by point splatting. (h) is the ground truth reference.  }
	\label{fig:eight}	
\end{figure*}

\noindent{\textbf{{3D Gaussian}}}
3D Gaussian is utilized for fine detail representation in the rendering stage.
To measure the effect of Gaussians, we replace Gaussian splatting with point splatting during rendering,
Each point is parameterized by the color $c$, the opacity $\sigma$ and the radius $r$ .
Fig. \ref{fig:eight} shows the difference between Gaussians and points on capturing fine details. 
The introduced point-based shape model can reconstruct the head including accessories like glasses, but struggles with the representation of single tiny hair. In a similar way, point splatting suffers from fine detail representation, thus causing the blurry hair in the green box of Fig. \ref{fig:eight}(e).
As a comparison, 3D Gaussian is considerable more flexible than points on 3D representation, and it can render sharp and realistic image.
%It is obvious that point splatting struggles with modeling the hair strand, and the image rendered by point splatting is blurry.
%As a comparison, Gaussian splats can represent the fine detail, ignored by points and can render realistic images.

\begin{table}[!t]	
	\caption{Ablation study on subject 1. "PS" refers to rendering with point splatting, 'GS' refers to rendering with Gaussian splatting, and 'ENH' means that the enhancement network is used. \colorbox[HTML]{E1EFDB}{Green} and \colorbox[HTML]{FFF2CE}{yellow} indicates the best and the second, respectively.}
	\label{table:two}
%	\begin{center}
	\centering
	\begin{tabular}{lccc}
		\hline\hline
		Subject ID   & \multicolumn{3}{c}{subject 1 (yufeng)} \\ \hline
		Metrics      & PSNR $\uparrow$        & SSIM  $\uparrow$      & LPIPS  $\downarrow$     \\ \hline
		w/o  $\mathcal{G}$          &25.7890      & 0.8951     & 0.0771     \\
		PS           & 28.4581      & 0.9016      & 0.0763      \\
		GS           & \cellcolor[HTML]{FFF2CE}29.1137      & 0.9126      & \cellcolor[HTML]{E1EFDB}0.0549      \\
		PS+ENH       
		& 29.0281     
		& \cellcolor[HTML]{FFF2CE}0.9183     
		& 0.0626    \\
		ours(GS+ENH) 
		& \cellcolor[HTML]{E1EFDB}29.3942     
		& \cellcolor[HTML]{E1EFDB}0.9212     
		& \cellcolor[HTML]{FFF2CE}0.0580      \\ \hline\hline
	\end{tabular}
%    \end{center}	
\end{table}
%\begin{figure}[!t]
%	\centering
%	\subfloat[]{
%		\begin{minipage}[b]{0.3\linewidth}		
%			\includegraphics[width=1\linewidth]{fig5-0.png}		
%		\end{minipage}
%	}
%	\subfloat[]{
%		\begin{minipage}[b]{0.3\linewidth}
%			\includegraphics[width=1\linewidth]{fig5-1.png}		
%		\end{minipage}
%	}
%	\subfloat[]{
%		\begin{minipage}[b]{0.3\linewidth}
%			\includegraphics[width=1\linewidth]{fig5-2.png}			
%		\end{minipage}
%	}	
%	\caption{The difference between point and 3D Gaussian on modeling shape. (a) shows the points trained with 2 epochs, (b) and (c) show the Gaussians trained with 2 and 10 epochs, respectively. It is clear that points converge quickly to capture the head shape.}	
%	\label{fig:six}	
%\end{figure}

\noindent{\textbf{{Enhancement}}} 
We have found that the U-net based enhancement can further improve the quality of the rendered image. 
As listed in Table \ref{table:two}, the enhancement network can quantitatively improve the visual quality in term of PSNR, SSIM and LPIPS.
Fig. \ref{fig:eight} indicates that the enhancement network can improve the visual quality of the rendered image.


\section{Conclusion}
We have developed PSAvatar, a novel framework for head avatar creation that facilitates flexible shape representation and efficient high-fidelity rendering.
PSAvatar utilizes a newly developed Point-based Shape Model (PSM) to reconstruct the surface-like geometry in the facial region and capture complex volumetric structures like glasses. The introduction of the PSM makes it possible to exploit the powerful and flexible representation capability of 3D Gaussian for fine detail representation and high-fidelity appearance modeling. We have shown that PSAvatar can create high-quality avatars of a variety of subjects and the avatars can be animated in real-time. Extensive experiments have demonstrated that PSAvatar has superior performances over strong baselines. 
%In addition, PSAvatar employs 3D Gaussian to further improve the 3D representation for fine details.
%Eventually, PSAvatar achieves head avatar creation on a variety of subjects.
%The reconstructed avatars show good performance on high-fidelity rendering and can be animated in real-time ($\ge$ 25fps at the resolution of 512 $\times$ 512) by the morphable model parameters.
%Comprehensive experiments demonstrate the superiority of the introduced method over the existing works.

{
    \small
    \bibliographystyle{ieeenat_fullname}
    \bibliography{main}
}

\end{document}